\documentclass[preprint]{aastex}
\usepackage{emulateapj5}

\newcommand{\sgrastar}{Sgr~A\mbox{$^*$}}
\newcommand{\msun}{\mbox{$\rm M_\sun$}}
\newcommand{\lsun}{\mbox{$\rm L_\sun$}}
\newcommand{\mdot}{\mbox{$\dot{M}$}}

\newcommand{\ledd}{\mbox{$L_{\rm E}$}}
\newcommand{\mdotedd}{\mbox{$\dot{M}_{\rm E}$}}
\newcommand{\chandra}{\emph{Chandra}}
\newcommand{\cxo}{\emph{CXO}}
\newcommand{\rosat}{\emph{ROSAT}}
\newcommand{\asca}{\emph{ASCA}}
\newcommand{\sax}{\emph{BeppoSAX}}
\newcommand{\ginga}{\emph{Ginga}}
\newcommand{\einstein}{\emph{Einstein}}

\newcommand{\granat}{\emph{Granat}}
\newcommand{\xmm}{\emph{XMM-Newton}}

\newcommand{\ariel}{\emph{Ariel~V}}
\newcommand{\hipparcos}{\emph{Hipparcos}}
\newcommand{\tycho}{\emph{Tycho-2}}
\newcommand{\usno}{\emph{USNO-A2.0}}

\slugcomment{Submitted to ApJ; 2001 February 2}
\shorttitle{\chandra\ Imaging of \sgrastar\ and the Galactic Center}
\shortauthors{Baganoff et al.}

\begin{document}

\title{\emph{Chandra} X-ray Spectroscopic Imaging of Sgr~A$^*$ \\ and
the Central Parsec of the Galaxy}

\author{F.~K.~Baganoff,\altaffilmark{1} Y.~Maeda,\altaffilmark{2}
        M.~Morris,\altaffilmark{3} M.~W.~Bautz,\altaffilmark{1}
        W.~N.~Brandt,\altaffilmark{2} W.~Cui,\altaffilmark{1,4}
        J.~P.~Doty,\altaffilmark{1} E.~D.~Feigelson,\altaffilmark{2}
        G.~P.~Garmire,\altaffilmark{2} S.~H.~Pravdo,\altaffilmark{5}
        G.~R.~Ricker,\altaffilmark{1} and L.~K.~Townsley\altaffilmark{2}}

%\email{fkb@space.mit.edu}

\altaffiltext{1}{Center for Space Research, Massachusetts Institute of
  Technology, Cambridge, MA 02139-4307; fkb@space.mit.edu}
\altaffiltext{2}{Department of Astronomy and Astrophysics, Pennsylvania
  State University, University Park, PA 16802-6305}
\altaffiltext{3}{Department of Physics and Astronomy, University of
  California, Los Angeles, CA 90095-1562}
\altaffiltext{4}{Department of Physics, Purdue University, West
Lafayette, IN 47907}
\altaffiltext{5}{Jet Propulsion Laboratory, California Institute of
  Technology, Pasadena, CA 91109}

\begin{abstract}

We present results of our \chandra\ observation with ACIS-I centered
on the position of Sagittarius~A$^{*}$ (\sgrastar), the compact
nonthermal radio source associated with the massive black hole (MBH)
at the dynamical center of the Milky Way Galaxy.  We have obtained the
first high spatial resolution ($\approx 1\arcsec$), hard X-ray
(0.5--7~keV) image of the central 40~pc (17\arcmin) of the Galaxy.

We have discovered an X-ray source, \objectname[]{CXOGC
J174540.0$-$290027}, coincident with the radio position of \sgrastar\
to within 0\farcs35, corresponding to a maximum projected distance of
16~light-days for an assumed distance to the center of the Galaxy of
8.0~kpc.  We received $222\pm17$ ($1\sigma$) net counts from the
source in 40.3~ks.  The source is detected with high significance,
$S/N \simeq 37\sigma$, despite the highly elevated diffuse X-ray
background in the central parsec of the Galaxy.  Due to the low number
of counts, the spectrum is well fit either by an absorbed power-law
model with photon index $\Gamma = 2.7^{+1.3}_{-0.9}$ ($N(E) \propto
E^{-\Gamma}$ photons cm$^{-2}$ s$^{-1}$ keV$^{-1}$) and column density
$N_{\rm H} = (9.8^{+4.4}_{-3.0}) \times 10^{22}$ cm$^{-2}$ (90\%
confidence interval) or by an absorbed optically thin thermal plasma
model with $kT = 1.9^{+0.9}_{-0.5}$~keV and $N_{\rm H} =
(11.5^{+4.4}_{-3.1}) \times 10^{22}$ cm$^{-2}$.  Using the power-law
model, the measured (absorbed) flux in the 2--10~keV band is
$(1.3^{+0.4}_{-0.2}) \times 10^{-13}$ ergs cm$^{-2}$ s$^{-1}$, and the
absorption-corrected luminosity is $(2.4^{+3.0}_{-0.6}) \times
10^{33}$ ergs s$^{-1}$.

The X-ray source coincident with \sgrastar\ is resolved, with an
apparent diameter of $\approx 1\arcsec$.  We report the possible
detection, at the $2.7\sigma$ significance level, of rapid continuum
variability on a timescale of several hours.  We also report the
possible detection of an Fe~K$\alpha$ line at the $\simeq 2\sigma$
level.  The long-term variability of \sgrastar\ is constrained via
comparison with the \rosat/PSPC observation in 1992.  The origin of
the X-ray emission (MBH vs.\ stellar) and the implications of our
observation for the various proposed MBH emission mechanisms are
discussed.  The current observations, while of limited
signal-to-noise, are consistent with the presence of both thermal and
nonthermal emission components in the \sgrastar\ spectrum.

We also briefly discuss the complex structure of the X-ray emission
from the Sgr~A radio complex and along the Galactic plane and present
morphological evidence that \sgrastar\ and Sgr~A West lie within the
hot plasma in the central cavity of Sgr~A East.  Over 150 point
sources are detected in the $17\arcmin \times 17\arcmin$ field of
view.  Our survey of X-ray sources is complete down to a limiting
2--10~keV absorbed flux of $F_{\rm X} \approx 1.7 \times 10^{-14}$
ergs cm$^{-2}$ s$^{-1}$.  For sources at the distance of the Galactic
Center, the corresponding absorption-corrected luminosity is $L_{\rm
X} \approx 2.5 \times 10^{32}$ ergs s$^{-1}$.  The complete
flux-limited sample contains 85 sources.  Finally, we present an
analysis of the integrated emission from the detected point sources
and the diffuse emission within the central 0.4~pc (10\arcsec) of the
Galaxy.

\end{abstract}


\section{Introduction}
\label{sec:intro}

After decades of controversy, measurements of stellar dynamics have
confidently established that the nucleus of the Milky Way Galaxy
harbors a massive black hole (MBH) with a mass $M = (2.6 \pm 0.2)
\times 10^{6}\ \msun$ \citep{Genzel97,Ghez98}.  The MBH coincides with
the compact, nonthermal radio source Sagittarius~A$^{*}$ (\sgrastar),
but no emission at other wavelengths has been convincingly associated
with it (\S\ref{sec:radio}).  It is also well known that the
bolometric luminosity ($L$) and the X-ray luminosity ($L_{\rm X}$) of
\sgrastar\ are far lower than expected from the standard thin
accretion disk model used in the study of X-ray binaries and quasars
\citep[][and references
therein]{Shakura73,Watson81,Bradt83,Frank92,Morris96}.  The bolometric
luminosity of a $2.6 \times 10^{6}\ \msun$ black hole radiating at the
Eddington rate ($\ledd$) is $\sim3 \times 10^{44}$ ergs s$^{-1}$,
while the measured bolometric luminosity of \sgrastar\ is $\la
10^{37}$ ergs s$^{-1}$ \citep[see][and references
therein]{Narayan98a}.  In the standard model, $\sim10\%$ of the
luminosity is in X-rays \citep{Frank92}, so one would expect $L_{\rm
X} \sim3 \times 10^{43}$ ergs s$^{-1}$, if \sgrastar\ were radiating
at the Eddington rate.  The MBH at \sgrastar\ has been undetected in
the 2--10~keV band with $L_{\rm X} < 10^{35}$ ergs s$^{-1}$
(\S\ref{sec:x-ray}), which is $\sim10^{9}$ times fainter than the
X-ray luminosity that would be expected at the Eddington rate.
Similarly, the $\sim10^{6}$ \msun\ MBHs at the cores of several nearby
spiral galaxies emit $L_{\rm X} \simeq 10^{37-39}$ ergs s$^{-1}$,
implying that they are $\sim 10^{5-7}$ times fainter in X-rays than
would be expected at their Eddington rates
\citep{Garcia00,Terashima00,Ho01}.

The absence of a strong, compact X-ray source associated with the MBH
at the Galactic Center has been one of the profound mysteries of
high-energy astrophysics and must have at least one of three basic
causes.  First, the MBH may reside in an environment where the
accretion rate $\mdot \ll \mdotedd = \ledd/(\eta c^2) \simeq 5.8
\times 10^{-3}\ \eta^{-1}$ \msun\ yr$^{-1}$, either because the
ambient gas has extremely low density, or because it is too hot or is
moving too fast to accrete efficiently, or because it is dynamically
ejected prior to accretion.  Here $\eta = L/(\mdot c^2)$ is the
radiative efficiency of the accretion flow and $c$ is the speed of
light.  Second, the mechanism of accretion may be such that the
radiative efficiency is extremely low.  The advection-dominated
accretion flow \citep[ADAF;][]{Narayan98b}\ and related models can
achieve low values of $\eta$ and have been intensively applied to the
\sgrastar\ problem.  Third, the X-ray emission from \sgrastar\ may be
much higher than observed due to anisotropy (e.g., a relativistic beam
oriented perpendicular to the Galactic plane) and/or extremely high
absorption along the line-of-sight.

The recently launched \emph{Chandra X-ray Observatory} (\cxo) with its
Advanced CCD Imaging Spectrometer (ACIS) detector provides a unique
opportunity to advance our knowledge of X-ray emission from \sgrastar.
It combines a superb mirror with sub-arcsecond resolution and an
imaging detector with high efficiency over a broad X-ray band and
moderate spectral resolution.  The spatial resolution and accurate
astrometry are essential to discriminate emission from \sgrastar\ from
X-rays produced in the surrounding compact cluster of massive stars
and other hot plasma in the region.  The sensitivity of \chandra/ACIS
at high X-ray energies is essential to penetrate the high interstellar
column density along the line-of-sight to the Galactic Center
(\S\ref{sec:radio}).

After a review of some relevant past studies
(\S\ref{sec:past_studies}), we describe the observations, data
analysis, source detection, and astrometry (\S\ref{sec:obs_anal}).
The resulting image of the inner 40~pc (17\arcmin) of the Galaxy is
presented in \S\ref{sec:gcimage}, and the properties of the innermost
arcsecond associated with \sgrastar\ are described in
\S\ref{sec:sgra_star}.  The integrated emission from point sources
(\S\ref{sec:pt_src_spec}) and the diffuse emission
(\S\ref{sec:diffuse_spec}) within the central 10\arcsec\ of the Galaxy
are discussed.  Tentative identifications of the X-ray point sources
in the central 10\arcsec\ with bright IR sources and the effects of
source confusion on observations by other X-ray satellites are
presented in \S\ref{sec:confusion}.  A limit on the long-term X-ray
variability of \sgrastar\ is derived in \S\ref{sec:long_term_var}.
The origin of the X-ray emission (i.e., MBH vs.\ stellar) is discussed
(\S\ref{sec:mbh_or_stellar}) and implications for the astrophysics of
the central MBH of the Galaxy are presented (\S\ref{sec:mbh_models}).
We summarize our findings, evaluate the various models, and discuss
the scientific goals of future observations in
\S\ref{sec:conclusions}.  This is the first of several papers arising
from this \chandra\ observation: our analysis of the X-ray emission
from Sgr~A East is presented by \citet{Maeda01}.  Future papers will
present our studies of the X-ray emission from the point sources and
the diffuse plasma distributed throughout the field.

\section{Past Studies}
\label{sec:past_studies}

\subsection{Radio/IR}
\label{sec:radio}

\sgrastar\ is a compact, nonthermal radio source
\citep{Balick74,Backer96}.  Radio proper motion studies performed over
the last decade place \sgrastar\ at the dynamical center of the Galaxy
and set a lower limit on its mass of $\ga 10^{3}\ \msun$
\citep{Backer99,Reid99}.  It has an intrinsic radio brightness
temperature $\ga 10^{10}$~K \citep{Backer93,Rogers94} and is weakly
variable on timescales of less than about a month in the centimeter
and millimeter bands \citep{Zhao89,Wright93,Falcke99,Zhao00}.  These
properties are reminiscent of the compact nuclear radio sources
present in radio-loud quasars and active galactic nuclei (AGN) and
suggest that \sgrastar\ may derive its luminosity from matter
accreting onto the MBH at the center of the Galaxy
\citep{Lynden-Bell71}.

Polarimetric and spectro-polarimetric observations made with the Very
Large Array (VLA) and the Berkeley-Illinois-Maryland-Association
(BIMA) radio interferometers show that \sgrastar\ is linearly
unpolarized at frequencies up to at least 86~GHz
\citep{Bower99a,Bower99c}; the upper-limit on linear polarization at
86~GHz is 1\%.  \citet{Aitken00}\ report the detection of linear
polarization from \sgrastar\ at 750, 850, 1350, and 2000~\micron\ with
the SCUBA camera on the 15-m James Clerk Maxwell Telescope (JCMT).
After removing the effects of strong free-free emission and polarized
dust from the single-dish JCMT beam (34\arcsec\ at 150 GHz),
\citeauthor{Aitken00}\ report the fractional linear polarization at
2000~\micron\ (150~GHz) to be $10^{+9}_{-4}\%$.

\citet{Bower99b}\ detect circular polarization of \sgrastar\ at 4.8
and 8.4~GHz with the VLA; the fractional circular polarization at
4.8~GHz is $-0.36 \pm 0.05\%$.  The circular polarization at 4.8~GHz
is confirmed independently by \citet{Sault99}\ with the Australia
Telescope Compact Array (ATCA).  \citet{Bower00}\ reviews the current
state of radio polarization observations of
\sgrastar.\footnote{GCNEWS: Galactic Center Electronic Newsletter
\citep{Falcke97}\ is available at
http://www.mpifr-bonn.mpg.de/gcnews/gcnews/Vol.11/article.shtml.}  He
reports significant variability of the circularly polarized flux on
timescales of a few days.  This has important implications for the
emission mechanism(s) at radio and (perhaps) higher frequencies.

The total radio luminosity of \sgrastar\ is estimated to be a few
hundred \lsun\ \citep{Morris96}.  This raises the possibility that the
emission could result from accretion onto a cluster of compact
stellar-mass objects \citep{Ozernoy89,Morris93}.  However, recent
proper motion studies of stars within 6\arcsec\ of the Galactic Center
constrain the minimum mass density of the central gravitational
potential to be $\ga10^{12}$ \msun\ pc$^{-3}$ \citep{Eckart97,
Ghez98}.  The best-fit model from \citeauthor{Ghez98}\ requires a dark
central object of mass $M = (2.6 \pm 0.2) \times 10^{6}\ \msun$ within
$\sim0.015$~pc of \sgrastar\ \citep[see also][]{Genzel00}.  These
results rule out a cluster of compact stellar mass objects as the
energy source for \sgrastar\ \citep[see][]{Maoz98} but provide no
direct evidence that the central engine is a MBH.  Furthermore,
dynamical studies cannot provide the spectral information needed to
identify the underlying emission mechanism or mechanisms.

Numerous models have been proposed which can produce centimeter
through millimeter band spectra that are at least roughly consistent
with the observations, but this spectral range is too narrow to
identify uniquely the nature of the central engine.  What is needed is
a detection or strict upper limit on the flux of \sgrastar\ at higher
frequencies to fix the overall spectrum on both ends.

A few claims have been made in the literature for the detection of
\sgrastar\ in the mid- and near-IR \citep{Stolovy96,Genzel97}.
However, the search for an infrared (IR) counterpart to \sgrastar\ is
hampered by source confusion and the strong IR background in the
Galactic Center.  Precise astrometric alignment of IR images with
radio maps using OH/IR stars that are also masers indicates that none
of the confirmed IR sources seen so far can be associated definitively
with the position of \sgrastar\ \citep{Menten97}.  Furthermore, none
of the near-IR sources yet stands out spectroscopically as a possibly
non-stellar object.  Consequently, claims of detection of IR emission
from \sgrastar\ are widely viewed as upper limits at this time.

The Galactic Center is heavily obscured by gas and dust in the optical
and ultraviolet wavebands \citep[$A_{\rm V} \sim
30$~mag;][]{Becklin78,Rieke89}.  Thirty magnitudes of visual
extinction corresponds to a column density $N_{\rm H} \approx 6
\times\ 10^{22}$ cm$^{-2}$ \citep{Predehl95}, so the obscuring medium
becomes partially transparent to X-rays from the Galactic Center at
energies $\ga 2$~keV.  X-ray observations thus provide our best
opportunity to constrain the high-frequency end of the spectral energy
distribution of \sgrastar.  Since strong, hard X-ray emission is a
characteristic property of AGN, \sgrastar\ is expected to be an X-ray
source if it derives its energy from accretion onto a massive black
hole.  However, no definitive detection of X-ray emission from
\sgrastar\ had been made prior to the launch of \chandra\ in July of
1999.

\subsection{X-ray}
\label{sec:x-ray}

X-ray observations of the regions surrounding \sgrastar\ were carried
out with very early rocket- and balloon-borne instruments \citep[see
review by][]{Skinner89}, but detailed observations started with
\einstein, the first satellite equipped with X-ray imaging optics
\citep{Watson81}.  \einstein\ observed the Galactic Center twice, 6
months apart, with the IPC (0.5--4.0~keV) for a total of 18.3~ks and
once with the HRI (0.5--4.5~keV) for 9.1~ks.  The \einstein/IPC images
have an angular resolution of $\sim 1\arcmin$ and reveal 12 discrete
sources within the central $1\arcdeg \times 1\arcdeg$ of the Galaxy.
The error box for the strongest of these sources (1E~1742.5$-$2859) is
centered only 20\arcsec\ from the position of \sgrastar.  Assuming an
absorbed thermal bremsstrahlung model with $kT = 5$~keV and $N_{\rm H}
= 6 \times 10^{22}$ cm$^{-2}$, \citeauthor{Watson81}\ estimate the
absorption-corrected 0.5--4.5~keV luminosity of this source to be $9.6
\times 10^{34}$ ergs s$^{-1}$.\footnote{Throughout this paper we adopt
8.0~kpc for the distance from Earth to the center of our Galaxy
\citep{Reid93}.  All luminosities have been adjusted to this distance,
except where specified otherwise.}  The \einstein\ images show that
the discrete sources are embedded in a bright, $25\arcmin \times
15\arcmin$ elliptically shaped region of apparently diffuse emission
lying along the Galactic plane that accounts for 85\% of all the
emission from that region.  No variability was detected in the point
sources over the 6-month baseline.  The \einstein/HRI image is
essentially blank due to the high absorbing column and the low
detection efficiency of that instrument.

\citet{Predehl94}\ observed the Galactic Center with the \rosat/PSPC
for 50~ks in March 1992 and detected 14 sources within the central
$30\arcmin \times 30\arcmin$ of the Galaxy.  With the relatively high
spatial resolution of 10--20\arcsec\ (FWHM), they resolved
1E~1742.5$-$2859 into three sources, one of which (RX~J1745.6$-$2900)
is coincident with the radio position of \sgrastar\ to within
10\arcsec.  The high absorption column and the soft energy band
(0.1--2.5~keV) of the PSPC limited their ability to fit the spectrum
and to constrain the spectral parameters.

Hard X-ray observations (2--30~keV) centered on \sgrastar\ have been
made with non-imaging instruments
\citep{Skinner87,Kawai88,Pavlinsky94}.  The line-of-sight to the
Galactic Center becomes optically thin to hard X-rays, hence the
fluxes measured by these missions are nearly free from the effects of
absorption.  These observations suggest the presence of a long-term
variable source near the position of \sgrastar\ with the measured
3--10~keV flux varying over the range (2--$10) \times 10^{-11}$ ergs
cm$^{-2}$ s$^{-1}$, corresponding to an unabsorbed luminosity of about
(2--$10) \times 10^{35}$ ergs s$^{-1}$.  In order to simultaneously
reproduce the low-energy spectrum measured with \rosat\ and a higher
energy spectrum (4--20~keV) measured with the ART-P telescope on
\granat\ \citep{Pavlinsky94}, \citeauthor{Predehl94}\ found the
absorption column has to be $N_{\rm H} \approx$ (1.5--$2) \times
10^{23}$ cm$^{-2}$, which is about 3 times larger than expected from
IR observations of nearby stars.  Following \citeauthor{Watson81},
\citeauthor{Predehl94} adopt a thermal bremsstrahlung model with $kT =
5$~keV, but $N_{\rm H} = 1.5 \times 10^{23}$ cm$^{-2}$, and derive an
unabsorbed 0.8--2.5~keV luminosity for RX~J1745.6$-$2900 of $6.6
\times 10^{35}$ ergs s$^{-1}$.\footnote{Using the same data as
\citet{Predehl94}, \citet{Predehl96} report the 0.8--2.5~keV
luminosity of RX~J1745.6$-$2900 as 1--$2 \times 10^{34}$ ergs
s$^{-1}$, assuming an absorbed power-law model with $\Gamma = 1.6$ and
$N_{\rm H} = 2 \times 10^{23}$ cm$^{-2}$.  However,
\citeauthor{Predehl96} do not specify whether this luminosity is
corrected for absorption.  We have used the spectral model of
\citeauthor{Predehl96} in XSPEC with the response matrix
\emph{pspcb\_gain2\_256.rsp} to compute the predicted PSPC count rate.
Normalizing the model to the count rate observed by
\citeauthor{Predehl94} ($8 \times 10^{-4}$ counts s$^{-1}$), we find
that the luminosity reported by \citeauthor{Predehl96} has not been
corrected for absorption.  Many papers in the literature use the
luminosity reported by \citeauthor{Predehl96} under the assumption
that it is corrected for absorption.  Consequently, the accretion
models in these papers, which are based in part on fits to the
luminosity reported by \citeauthor{Predehl96}, underestimate the
upper-limits on the accretion rate and the X-ray luminosity of
\sgrastar\ in 1992 by 1--2 orders of magnitude.\label{foot:rosat}}

The first hard X-ray imaging (up to 10~keV) with modest spatial
resolution (3\arcmin) was made in 1993 with \asca\ \citep{Koyama96}.
\asca\ detected diffuse thermal emission ($kT \approx 10$~keV) with
helium-like and hydrogen-like K$\alpha$ emission lines of various
elements covering the central 1~square degree of the Galactic Center.
A $2\arcmin \times 3\arcmin$ elliptical region that fills the Sgr~A
East shell showed bright diffuse emission at a level 5 times that of
the more extended emission.  After correcting for a measured
absorption of $N_{\rm H} \approx 7 \times 10^{22}$ cm$^{-2}$, the
unabsorbed 2--10~keV luminosity of this gas was found to be
$\sim10^{36}$ ergs s$^{-1}$.  No subtraction was performed for the
variable local background, consequently \asca\ could only place an
upper-limit of $\sim10^{36}$ ergs s$^{-1}$ on the X-ray luminosity of
\sgrastar.

\citet{Koyama96}\ found a hard X-ray source located $1\farcm3$ away
from \sgrastar.  During their second observation made in 1994,
\citet{Maeda96}\ discovered an X-ray burst and eclipses with a period
of 8.4~hr from the hard source, establishing that it is an eclipsing
low-mass X-ray binary.  Only one cataloged transient source,
A~1742$-$289 \citep{Eyles75}, which appeared in 1975, positionally
coincides within the error region.  However, \citet{Kennea96}\
reanalyzed \ariel\ data taken in 1975 and found no eclipses from
A~1742$-$289.  Hence the hard source was identified as a newly
discovered low-mass X-ray binary and was named AX~J1745.6$-$2901.
\citet{Maeda96}\ report that the absorbed flux from AX~J1745.6$-$2901
varied from $1 \times 10^{-11}$ to $4 \times 10^{-11}$ ergs cm$^{-2}$
s$^{-1}$, which is similar to the range of variability of the X-ray
flux from the \sgrastar\ region previously reported by the non-imaging
instruments.  These results indicate that the fluxes measured with the
non-imaging instruments and attributed to \sgrastar\ may be
contaminated significantly by AX~J1745.6$-$2901 and possibly by
A~1742$-$289.

A \sax/MECS observation, with about 1\farcm3 angular resolution on
axis and an energy range similar to the {\asca}/SIS, was performed in
1997, 4 years after the first {\asca} observation \citep{Sidoli99}.
\sax\ detected the diffuse emission near \sgrastar, measured the
absorption column to be $N_{\rm H} \approx 8 \times 10^{22}$
cm$^{-2}$, and set a tighter upper limit on the 2--10~keV luminosity
of \sgrastar\ of $L_{\rm X} \la 10^{35}$ ergs s$^{-1}$.  No indication
of a hard X-ray counterpart to the MBH at \sgrastar\ was found with
\asca\ or \sax.

\section{Observations and Analysis}
\label{sec:obs_anal}

\subsection{Data Acquisition and Reduction}
\label{sec:reduce}

\chandra\ \citep{Weisskopf96} observed the center of our Galaxy with
ACIS-I, the imaging array of the Advanced CCD Imaging Spectrometer
(G.~P.~Garmire, J.~P.~Nousek, \& M.~W.~Bautz, in preparation), for
51.1~ks on 1999 September 21.  All four CCDs (I0--3) in the $2 \times
2$ imaging array (ACIS-I) and the central two CCDs (S2--3) in the $1
\times 6$ spectroscopy array (ACIS-S) were used.  The photo-sensitive
region of each CCD is comprised of $1024 \times 1024$ pixels, with
each pixel subtending $0\farcs5 \times 0\farcs5$ on the sky; hence,
each CCD subtends $8\farcm3 \times 8\farcm3$ on the sky.  Detector S3
is a backside-illuminated CCD, while the other five are
frontside-illuminated.  The ACIS CCDs were clocked in timed-exposure
(TE) mode using the standard integration time of 3.2~s per frame.  The
focal plane temperature was $\approx -110\,^\circ$C.  To prevent
telemetry saturation, events with energies $\ga 15$~keV and events
with ACIS flight grade 24, 66, 106, 214, or 512 were rejected on
orbit.\footnote{See \S6.3 of the Chandra Proposer's Observatory Guide
Rev.\ 2.0, hereafter the POG, for the definition of the ACIS flight
grades and their correspondence with \asca\ grades.}  The data were
telemetered to the ground in very-faint (VF) mode; in this mode the
telemetered data contains the pulse-height amplitudes (PHAs) of a $5
\times 5$-pixel island centered on each event.

During ground processing, we further rejected events with \asca\ grade
1, 5, or 7 and events with certain ACIS flight grades located on CCD
quadrant boundaries to minimize the quiescent instrumental background.
Additional filtering was performed to exclude periods of time during
which large background flares saturated telemetry, causing the
majority of frames in these intervals to be lost.  The total exposure
time after filtering was 37.3~ks for S3 and 40.3~ks for each of the
frontside-illuminated chips.

Analyses of on-orbit data by the Chandra X-ray Center (CXC) and others
have shown that the frontside-illuminated CCDs occasionally exhibit
``flaring pixels''.  This phenomenon occurs when a cosmic ray deposits
a large amount of charge in traps at the interface between the active
region and the insulating layer of the gate structure.\footnote{A
description of the ``flaring-pixel'' or ``cosmic-ray afterglow''
problem in the ACIS frontside-illuminated CCDs is available from the
CXC at http://asc.harvard.edu/ciao/caveats/acis.html.}  The
de-trapping time constant is longer than the integration time between
frames, so events with identical grades are reported in the same pixel
in up to 7 consecutive frames.  These events can thus appear as false
sources with $\la 7$ counts.  The CXC has developed an algorithm
(\emph{acis\_detect\_afterglow}) for removing flaring pixel events.
However, the current algorithm removes a significant fraction of the
events from real sources as well.\footnote{See the note from the ACIS
Instrument Team posted at
http://asc.harvard.edu/ciao/caveats/acis\_cray.html.}  We have
examined our source list (\S\ref{sec:astrom}) and find that about 10\%
of the 158 sources have fewer than 8 counts, while 85\% have at least
10 counts.  Given the small number of potential flaring-pixel events,
the expected number of flaring pixels that overlap with a real X-ray
source within say 3\arcsec\ is $\la 0.06$.  The key results of this
paper (i.e., the astrometry and the spectral analyses) are based on
sources with $\gg 10$ counts, so these results will not be affected by
flaring pixels.  We have therefore chosen not to filter the data for
flaring pixels at this time.

Early in the \chandra\ mission, the frontside-illuminated CCDs
suffered radiation damage believed to be caused by low-energy protons
scattering off the high-resolution mirror assembly (HRMA) during
repeated passages through the Earth's radiation belts
\citep{Prigozhin00}.  This radiation introduced charge traps in the
buried channels of the CCDs that increased their charge-transfer
inefficiency (CTI).  At the focal plane temperature of
$-110\,^\circ$C, the integrated spectrum of the five
frontside-illuminated CCDs cuts off rapidly below $\approx 0.5$~keV
due to the increased CTI.  In addition, the instrumental background
begins to dominate the spectrum for energies $\ga7$--8~keV
\citep[e.g.,][and this paper]{Baganoff99}.  Therefore, the maximum
signal-to-noise ratio for the integrated spectrum occurs in the energy
range from about 0.5 to 7~keV.

No attempt was made to correct the observed flight event grades for
grade migration caused by the increased CTI.  Event amplitudes were
computed using the PHAs from the central $3 \times 3$-pixels of each
event.  The event amplitudes were converted to energies using the
routine \emph{acis\_process\_events} with the standard gain
file\footnote{acisD1999-09-16gainN0003.fits} provided by the CXC for
in-flight data taken at $-110\,^\circ$C.  The {\em
acis\_process\_events} routine is part of the Chandra Interactive
Analysis of Observations (CIAO) software package developed by the CXC.
Data reduction was performed using CIAO~1.1.3.

\subsection{Source Detection and Astrometry}
\label{sec:astrom}

We ran \emph{wavdetect}, the CIAO wavelet source detection routine
\citep{Dobrzycki99,Freeman01}, on an image formed from events in the
0.5--7~keV band using kernel scales ranging from 1 to 16 pixels in
multiples of $\sqrt2$; each ACIS pixel is 0\farcs5 on a side. The
source significance threshold was set equal to $1 \times 10^{-7}$;
since each ACIS CCD has about $1 \times 10^{6}$ pixels, the expected
number of false detections in all six CCDs is $\approx0.6$.

\emph{Wavdetect} found 158 sources: 157 sources in I0--3, 1 source in
S2, and 0 sources in S3.  The deficiency of sources in S2 and S3 is
attributable to a combination of the mirror vignetting, the enlarged
point-spread function (PSF) far off-axis ($\ga10\farcm3$), and the
two-times higher background rate in S3 compared to the
frontside-illuminated devices.  A detailed study of the point sources
in the field, including the $\log N$--$\log S$ distribution, will be
presented in another paper.  Here we give just a few preliminary
results.  A histogram plot of the cumulative number of sources
detected with greater than or equal to $C$ counts $N(\ge C)$ in the
0.5--7~keV band shows that our survey of X-ray sources is complete
down to $\approx 20$ counts or $\approx 5.0 \times 10^{-4}$ counts
s$^{-1}$.  To convert the 0.5--7~keV count rates to absorbed fluxes
($F_{\rm X}$) and absorption-corrected luminosities ($L_{\rm X}$) in
the 2--10~keV band, we adopt an absorbed Crab-like spectrum with
photon index $\Gamma = 2$ and column density $N_{\rm H} = 1 \times
10^{23}$ cm$^{-2}$ (see \S\ref{sec:sgra_star_continuum}).\footnote{All
of the X-ray luminosities ($L_{\rm X}$) presented in this paper are
corrected for absorption, while the fluxes ($F_{\rm X}$) are not
corrected for absorption.}  The corresponding completeness limit is
$F_{\rm X} \approx 1.7 \times 10^{-14}$ ergs cm$^{-2}$ s$^{-1}$.  For
sources at the distance of the Galactic Center, the corresponding
luminosity is $L_{\rm X} \approx 2.5 \times 10^{32}$ ergs s$^{-1}$.
Excluding the source in S2, the complete flux-limited sample contains
85 sources; hence, the mean X-ray source density of the complete
flux-limited sample within the $17\arcmin \times 17\arcmin$ ACIS-I
field of view (FOV) is $0.30 \pm 0.03$ sources arcmin$^{-2}$, and the
mean density of all the sources within the ACIS-I FOV is $0.54 \pm
0.04$ sources arcmin$^{-2}$.  Note that this preliminary analysis
ignores the variations in the absorbing column and the off-axis PSF
across the field.  For comparison, the on-axis $3\sigma$ detection
limit is $F_{\rm X} \approx 2.4 \times 10^{-15}$ ergs cm$^{-2}$
s$^{-1}$, and the corresponding luminosity is $L_{\rm X} \approx 3.5
\times 10^{31}$ ergs s$^{-1}$.

The CXC has measured the on-orbit performance of the Pointing Control
and Aspect Determination (PCAD) system on \chandra\ (see \S5.4 and
Table~5.1 of the POG).  Their analysis shows that standard CXC
processing is capable of placing a reconstructed X-ray image on the
celestial sphere to an accuracy of $0\farcs76$ (RMS) radius.  This
corresponds to a projected distance of about 0.03~pc or 35~light-days
at the Galactic Center.  To improve on this, we used sources in the
\tycho\ optical astrometric catalog from the \hipparcos\ satellite
\citep{Hog00} to register the \chandra\ field on the sky.

The center of our Galaxy is highly obscured (\S\ref{sec:radio}), so
known optical sources in the field of view must be relatively near the
Earth.  The obscuring medium becomes partially transparent to X-rays
from the Galactic Center at energies above about 2~keV.  Therefore, we
ran \emph{wavdetect} on a 0.5--1.5~keV image to select foreground
X-ray sources.  This yielded a total of 72 foreground sources in the
field of view: 71 sources in I0--3 and 1 source in S2.

\begin{figure}[ht]
%\epsscale{1.0}
\plotone{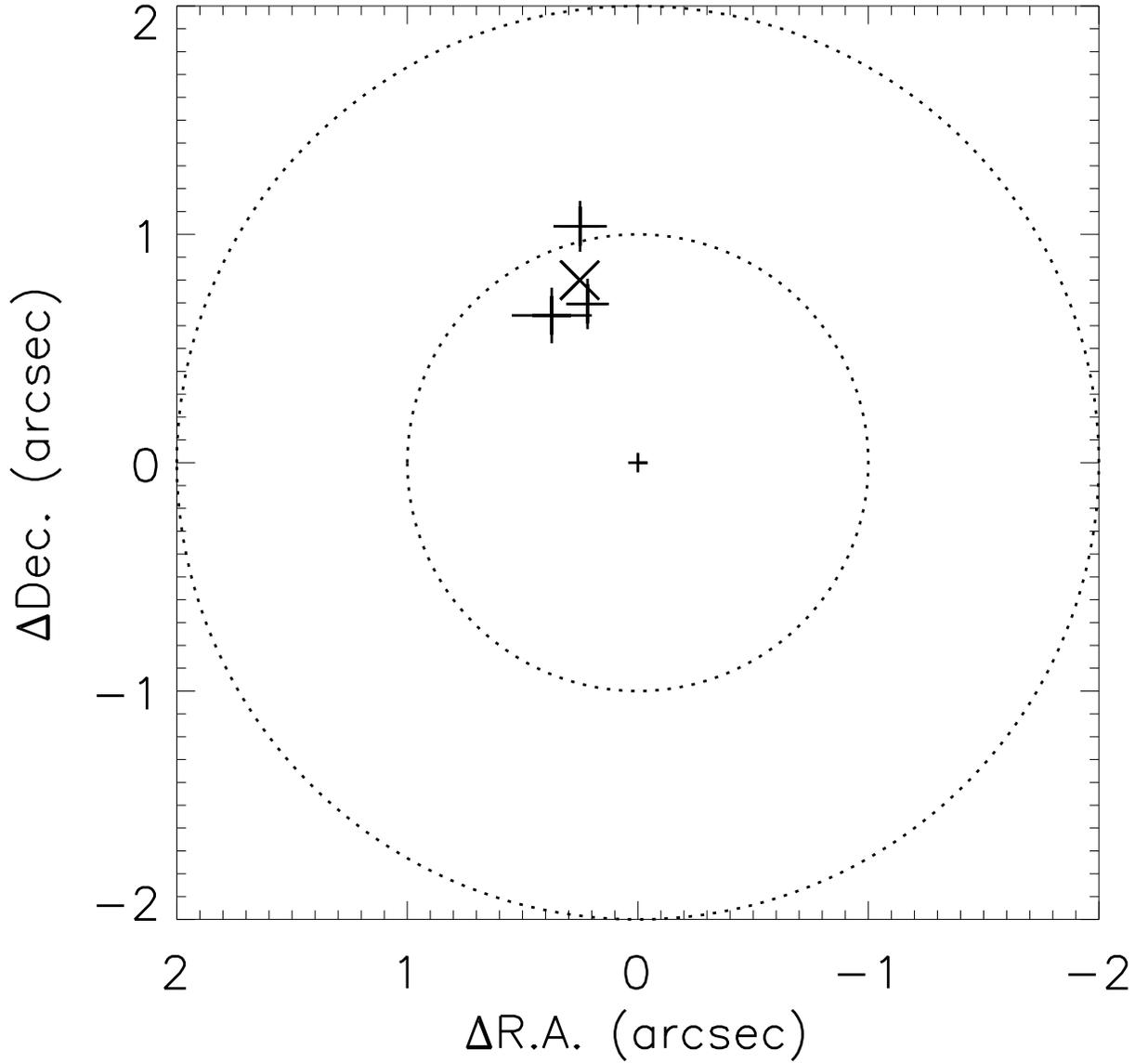}
\caption{Astrometric reference source positional offsets (\tycho\ $-$
\chandra) for matches within a 2\arcsec\ correlation radius
(\S\ref{sec:astrom}).  Error bars are $1\sigma$ formal uncertainties.
The small symbol at the origin indicates the relative position of the
\chandra\ boresight on the sky.  The cross at ($+0\farcs25$,
$+0\farcs80$) marks the weighted mean offset of the reference sources.
This offset was added to the celestial coordinates of the \chandra\
boresight to register the field on the \hipparcos\ celestial
coordinate system.
\label{fig:tycho_offset}}
\end{figure}

To minimize any potential effects of the variable off-axis PSF on
source centroids, we restricted the search to sources within 7\arcmin\
of the telescope boresight.  This region contains 7 \tycho\ sources
and 50 \chandra\ sources in the 0.5--1.5~keV band.  We found 3 matches
using a correlation radius of 2\arcsec\ (see
Table~\ref{tab:tycho_astrom}).  The expected number of false matches
is $7.9 \times 10^{-3}$; this quantity is equivalent to the cumulative
probability of getting at least 1 false match.  The probability of
getting 3 matches out of 7 trials by random chance is $5.0 \times
10^{-8}$.  It is therefore highly likely that all 3 matches are real.
The astrometric uncertainties listed in the \tycho\ catalog for the
positions and proper motions of the 3 reference stars range from
25--104~mas and 1.8--4.3~mas yr$^{-1}$, respectively.  The source
offsets (\tycho\ position $-$ \chandra\ position) shown in
Figure~\ref{fig:tycho_offset} are all in close agreement, indicating
that the celestial location of the boresight should be adjusted
slightly east and north to align the \chandra\ field to the
\hipparcos\ celestial coordinate system.  The weighted mean offset of
the reference sources is $\Delta\alpha = +0\farcs25$ and $\Delta\delta
= +0\farcs80$.  An independent check using 11 matching sources from
the \usno\ catalog \citep{Monet98} is consistent with the \tycho\
offset to within 0\farcs40.  This level of disagreement is consistent
with the larger uncertainties in the \usno\ positions ($\approx
0\farcs25$), which were measured from optical plates, and the lack of
correction for proper motion in the \usno\ positions.

We applied the weighted mean offset given above to register the
\chandra\ field on the sky and to correct the celestial locations of
all the X-ray sources.  The registered boresight is located 15\farcs8
east and 11\farcs0 south of the radio position of \sgrastar.  The
residual RMS scatter in the corrected X-ray positions of the \tycho\
reference sources is 0\farcs23; hence the astrometric uncertainty of
the registered field is 0\farcs13.

\section{X-ray Images of the Galactic Center}
\label{sec:gcimage}

We generated a raw, broad-band \chandra\ image of the center of our
Galaxy by binning 0.5--7~keV counts from the event list into a
two-dimensional image.  The resulting image suffered from the effects
of the mirror vignetting and the gaps between the CCDs.  In addition,
it was difficult to see low-surface-brightness extended emission.  We
developed a method for smoothing and flat-fielding the raw image to
remove these effects.  We describe our method here.

The mirror vignetting and the effective area curve for the combined
HRMA/ACIS instrument are both energy dependent (see Figs.~4.3 and 6.9
of the POG).  We split the broad 0.5--7~keV band into several
narrower bands to minimize variation of the effective area across
each band, while at the same time creating images with a reasonable
number of counts in astrophysically interesting bands.  Based on these
criteria, we create narrow-band images in the 0.5--1.5, 1.5--3, 3--6,
and 6--7~keV bands.  We then use the CIAO routines
\emph{mkinstmap} and {\em mkexpmap} to create monochromatic exposure
maps at 1, 2.4, 5, and 6.4~keV; these energies are selected because
the effective area at each energy roughly approximates the mean
effective area over the corresponding band, after allowing for the
characteristically steep spectral shape of Galactic X-ray sources and
the large column density toward the Galactic Center.  The narrow-band
images and the corresponding exposure maps are then binned using $6
\times 6$-pixel (i.e., $3\arcsec \times 3\arcsec$) bins to increase
the chances of getting at least 1 count per bin within the chip-gap
regions and to cut down on the computational time in subsequent steps.

Next we run the CIAO routine \emph{csmooth} to adaptively smooth the
broad-band image using minimum and maximum signal-to-noise thresholds
of 3$\sigma$ and $5\sigma$, respectively.  The background level is
computed locally.  The \emph{csmooth} routine is based on the method
of \citet{Ebeling01}.  One of the outputs generated by \emph{csmooth}
is a scale map recording the size of the gaussian smoothing kernel
used at each point in an image.  This scale map is input back into
\emph{csmooth} as we smooth each narrow-band image and exposure map so
that all images and maps are adaptively smoothed in exactly the same
way.\footnote{The present version of \emph{csmooth} fails to conserve
flux precisely (M.~Machacek 2000, private communication).  While this
problem does not significantly affect the results presented here, it
will be addressed in our subsequent paper on the diffuse emission.}
After smoothing, we divide each image by its corresponding exposure
map to produce a flat-fielded narrow-band image; these narrow-band
images are then added together to produce a flat-fielded broad-band
image.

\begin{figure}[ht]
%\epsscale{1.0}
\plotone{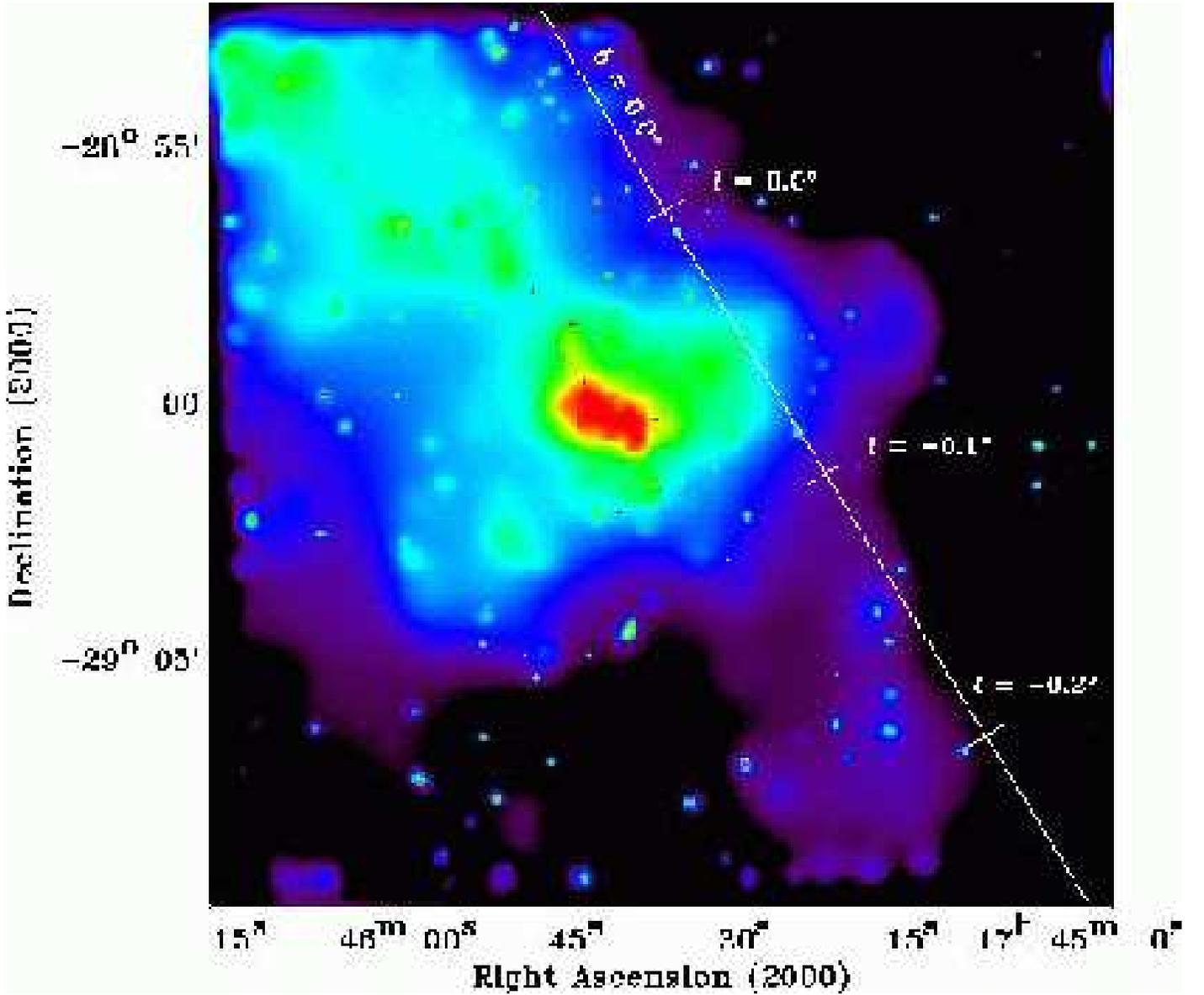}
\caption{\chandra\ 0.5--7~keV image of the central $17\arcmin \times
17\arcmin$ of the Galaxy (\S\ref{sec:gcimage}).  The image has been
adaptively smoothed and flat-fielded to bring out the low surface
brightness emission and to remove the effects of the mirror vignetting
and the gaps between the CCDs.  The red region at the center fills the
nonthermal, shell-like radio source Sgr~A East and sits on a ridge of
X-ray emission (green and blue) extending north and east parallel to
the Galactic plane (white line).  Also visible are regions of diffuse
emission extending perpendicular to the plane through the position of
\sgrastar\ (see
Figs.~\ref{fig:sgra_east}--\ref{fig:raw_central_arcmin}) and many of
the over 150 point sources detected in the field.  This image and the
other images in this paper are uncorrected for the effects of the
variations in the column density and the HRMA PSF across the field.
\label{fig:gal_center}}
\end{figure}

\begin{figure}[ht]
%\epsscale{1.0}
\plotone{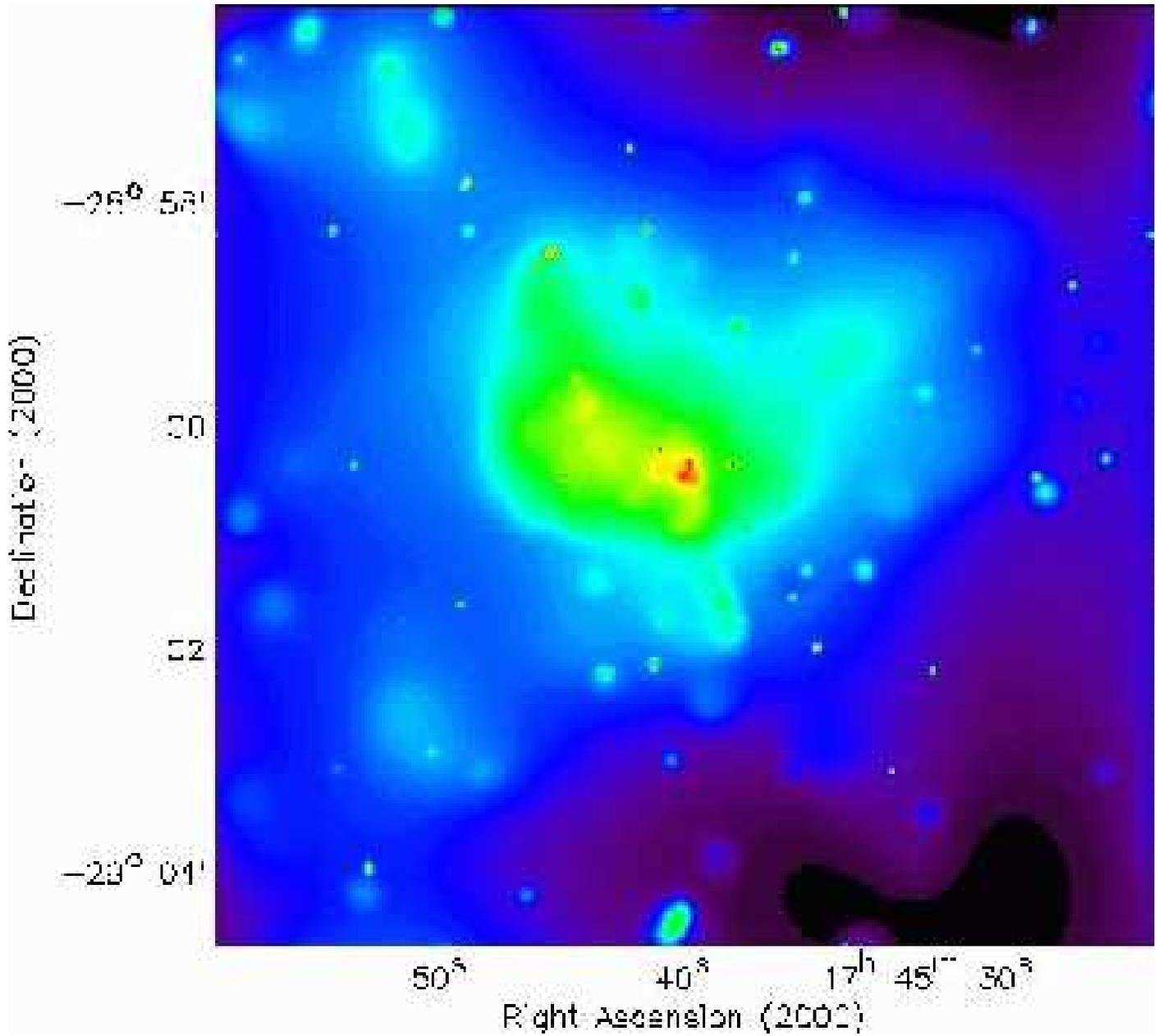}
\caption{\chandra\ 0.5--7~keV image of the central $8\farcm4 \times
8\farcm4$ of the Galaxy (\S\ref{sec:gcimage}), showing structure in
the X-ray emission from the vicinity of Sgr~A East (yellow and
green).  The image has been adaptively smoothed and flat-fielded.
X-ray emission from the compact, nonthermal radio source \sgrastar\ is
just discernable as the southeastern component of the red structure at
the center of the image. \label{fig:sgra_east}}
\end{figure}

\begin{figure}[ht]
%\epsscale{1.0}
\plotone{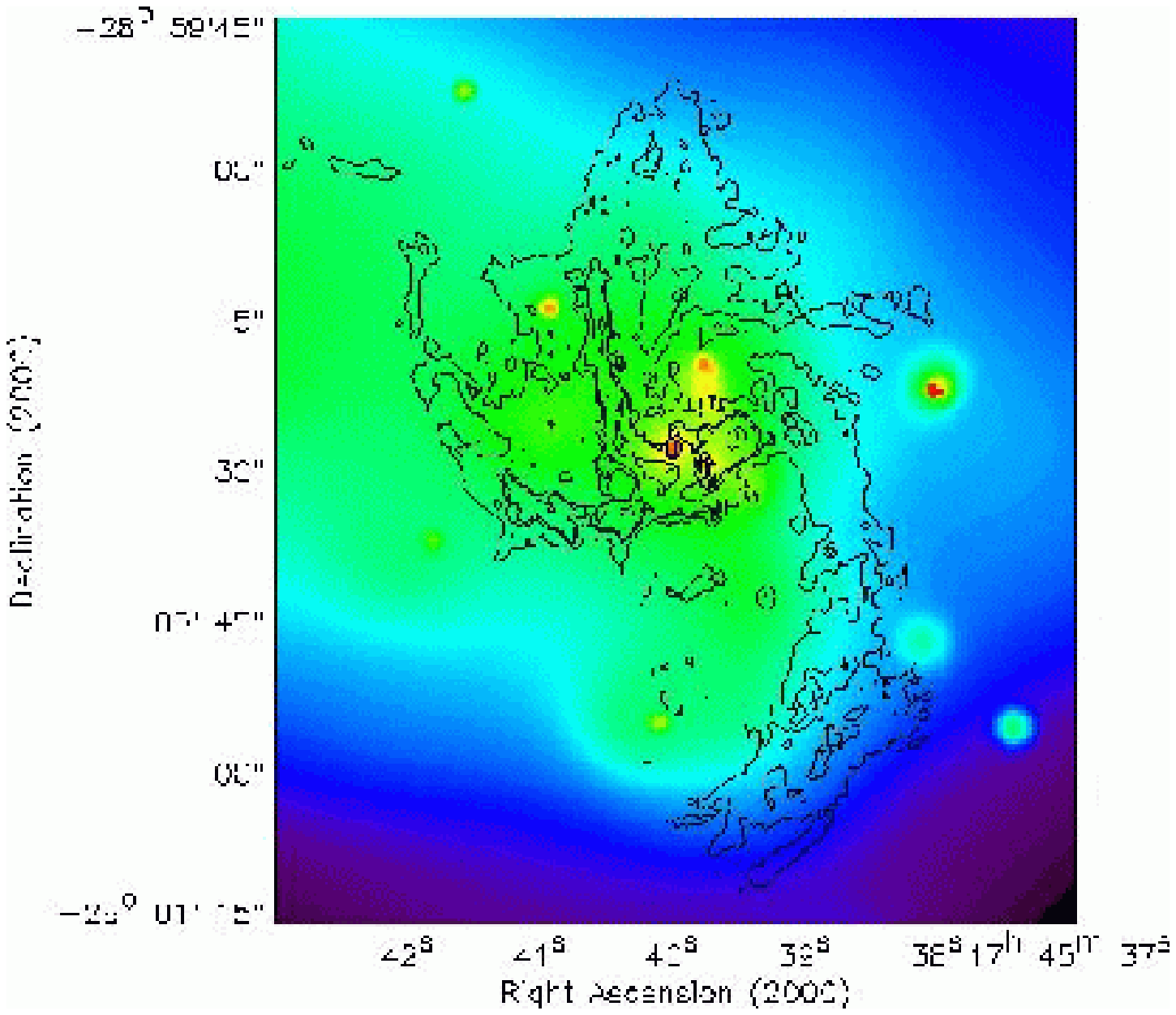}
\caption{\chandra\ 0.5--7~keV image of the central $1\farcm3 \times
1\farcm5$ of the Galaxy (\S\ref{sec:gcimage}).  The image has been
adaptively smoothed and flat-fielded.  Overlaid on the image are VLA
6-cm contours of \sgrastar\ and Sgr~A West from F.~Yusef-Zadeh (1999,
private communication).  X-ray emission from the vicinity of
\sgrastar\ appears as a red dot at $\rm 17^h45^m40.0^s$,
$-29\arcdeg00\arcmin28\arcsec$.  X-ray emission coincident with IRS~13
(yellow) is evident just southwest of
\sgrastar. \label{fig:sgra_west}}
\end{figure}

In Figures~\ref{fig:gal_center}--\ref{fig:sgra_west}, we present an
exploded view of the center of our Galaxy made with \chandra/ACIS-I in
the 0.5--7~keV band.  These images have been adaptively smoothed
and flat-fielded as described above.  Remarkable structure in the
X-ray emission from the Galactic Center is revealed for the first time
with sufficient angular resolution to allow detailed comparisons with
features seen in the radio and IR wavebands.

Figure~\ref{fig:gal_center}\ is a false-color image of the full
$17\arcmin \times 17\arcmin$ ACIS-I field of view, covering the
central 40~pc of the Galaxy.  The Galactic plane is marked by a white
line.  Numerous point sources and bright, complex diffuse emission are
readily visible.  The diffuse X-ray emission is strongest in the
center of Sgr~A East (red region), a well-known, shell-like,
nonthermal radio source.  The origin of Sgr~A East has been a topic of
debate since its discovery.  It has been interpreted by some as a
supernova remnant \citep[SNR;][]{Jones74,Ekers83}, but alternative
origins have been proposed as well
\citep[e.g.,][]{Yusef-Zadeh87,Mezger89,Khokhlov96}.  Our detailed
study of the X-ray counterpart \citep{Maeda01}\ argues strongly that
Sgr~A East is a rare type of metal-rich, ``mixed-morphology'' (MM)
supernova remnant that may have been produced about 10,000 years ago
by the Type~II explosion of a 13--$20\ \msun$ progenitor.  The X-ray
emission from Sgr~A East is concentrated in the central 2--3~pc within
the $6 \times 9$~pc radio shell and offset about 2~pc from \sgrastar.
The spectrum shows a thermal plasma ($kT \simeq 2$~keV) with strongly
enhanced metal abundances and elemental stratification.

A curious linear feature $\sim 0\farcm5$ long (yellow) extends (in
projection) from the northeast toward the center of Sgr~A East.  Its
appearance resembles that of a ``plume'' of emission sticking out the
top of Sgr~A East.  The brightest part of the \objectname[]{Sgr~A
``plume''} is located at $\rm 17^h45^m44\fs5$,
$-28\arcdeg59\arcmin36\arcsec$ (J2000.0).  It is clearly present in
raw narrow-band images in the 1.5--3 and 3--6~keV bands, but it is not
visible in the 6--7~keV band.  This is in contrast to the core X-ray
emission from Sgr~A East, which dominates the 6--7~keV band due to
strong Fe-K$\alpha$ line emission.

Sgr~A East sits on a ridge of emission (green \& blue) extending
north and east parallel to the Galactic plane that was first seen by
\einstein\ \citep{Watson81} and later observed by \rosat\
\citep{Predehl94}, \asca\ \citep{Koyama96}, and \sax\
\citep{Sidoli99}.  This ridge is most sharply defined in the 3--6~keV
band, with clumps of bright emission visible in the 6--7~keV band.
Spectral analysis of the \asca\ data by \citeauthor{Koyama96}
indicated that this emission is from a thermal plasma with $kT \approx
10$~keV, but our preliminary analysis of the \chandra\ data suggests
the emission is from a much cooler gas ($kT \approx 3$~keV).

Emission (green \& blue) extending perpendicular to the Galactic plane
in both directions through the position of \sgrastar\ is clearly
visible for the first time.  This extended X-ray emission appears to
correspond spatially with the so-called Sgr~A ``halo'' in the radio
band, but further study will be required to determine whether or not
there is a detailed correlation.  The emission is strong in the
1.5--6~keV band but absent at 6--7~keV.  We note that a line drawn
between the centers of the two brightest regions of this structure
(see also Fig.~\ref{fig:sgra_east}) would run directly through the
position of \sgrastar; this may indicate the presence of some sort of
hot, ``bipolar'' outflow from the vicinity of the MBH.  In that case,
the X-ray emitting plasma may be escaping preferentially along
magnetic field lines, which, at the center of the Galaxy, run
perpendicular to the Galactic plane to within about 20\arcdeg\
\citep{Morris85,Anantharamaiah91,Morris96,Lang99}.

Figure~\ref{fig:sgra_east}\ is an expanded view centered on \sgrastar\
of the inner $8\farcm4 \times 8\farcm4$ of the field.  This image was
created using the procedure described above starting with $2 \times
2$-pixel bins.  Complex structures can be seen in the X-ray emission
from the vicinity of Sgr~A East (yellow and green).  The ``plume''
discussed above in Figure~\ref{fig:sgra_east}\ is aligned with a
string of clumps or knots (yellow) within Sgr~A East, implying that
this feature might in fact be physically related to Sgr~A East rather
than simply a chance superposition on the sky.  X-ray emission that we
associate with the compact, nonthermal radio source \sgrastar\ is just
discernable in this panel as the southeastern component of the red
structure at the center of the image.  In addition, there is a clump
of bright emission (yellow) centered $\approx0.3$~pc east of
\sgrastar.

Figure~\ref{fig:sgra_west}\ is a $1\farcm3 \times 1\farcm5$ close-up
around \sgrastar\ (red dot at $\rm 17^h45^m40\fs0$,
$-29\arcdeg00\arcmin28\arcsec$ J2000.0) overlaid with VLA 6-cm
contours of \sgrastar\ and Sgr~A West (F.~Yusef-Zadeh 1999, private
communication).  The image was created as described above starting
with a full-resolution (i.e., $1 \times 1$-pixel binned) image.
Sgr~A West is an \ion{H}{2}\ region seen in absorption against the
nonthermal emission from Sgr~A East; consequently, Sgr~A West must
lie in front of Sgr~A East \citep{Yusef-Zadeh87,Pedlar89}.  The
absorption is not total, however, so Sgr~A West may lie near the
front edge of the Sgr~A East shell.  X-ray emission coincident with
\objectname[]{IRS~13} (yellow) is evident just southwest of \sgrastar.

The western boundary of the brightest diffuse X-ray emission (green)
coincides precisely with the shape of the Western Arc of the thermal
radio source Sgr~A West.  On the eastern side, the emission continues
smoothly into the heart of Sgr~A East (see red region in
Fig.~\ref{fig:gal_center}).  In addition, the indentation seen in the
X-ray intensity $\approx 25\arcsec$ southeast of \sgrastar\ coincides
with a molecular emission peak in the circumnuclear disk
\citep[CND;][]{Wright87,Marr93,Yusef-Zadeh00,Wright01}.  Since the
Western Arc is believed to be the ionized inner edge of the CND, the
morphological similarities between the X-ray and the radio structures
strongly suggest that the brightest X-ray-emitting plasma is being
confined by the western side of the CND.  This may be evidence that
Sgr~A West and \sgrastar\ physically lie within the hot cavity inside
the Sgr~A East shell.  We discuss this possibility further in our
companion paper on the X-ray emission from Sgr~A East
\citep{Maeda01}.  The alternative possibility, that Sgr~A East and
West occupy physically separate regions of space, requires a chance
alignment of the CND along our line of sight to the western edge of
Sgr~A East.  The morphological similarities would then be simply the
result of obscuration by the molecular gas and dust in the CND.

We are using the narrow-band images described above to study the
distribution of hard and soft point sources in the field and to study
the morphology of the bright Fe~K$\alpha$-line emission first observed
by \ginga\ \citep{Koyama89}.  These results will be presented
elsewhere.

\section{X-ray Emission from the Position of \sgrastar}
\label{sec:sgra_star}

\subsection{Position}
\label{sec:sgra_star_position}

Figure~\ref{fig:raw_central_arcmin} shows a $1\arcmin \times 1\arcmin$
field that is centered on the position of \sgrastar\ and is made from
the counts in the 0.5--7~keV band.  This image has not been
smoothed or flat-fielded.  The black cross marks the radio
interferometric position of \sgrastar\ as determined by
\citet{Yusef-Zadeh99}.  Clearly visible at the center of the image is
the X-ray source, \objectname[]{CXOGC J174540.0$-$290027}, that we
associate with \sgrastar\ based on the extremely close positional
coincidence.  The \emph{wavdetect} centroid position of
\objectname[]{CXOGC J174540.0$-$290027} is offset 0\farcs35 from the
radio position of \sgrastar, corresponding to a maximum projected
distance of 16~light-days (see Table~\ref{tab:sgra_star_astrom}).  The
uncertainty in the position of \objectname[]{CXOGC J174540.0$-$290027}
is 0\farcs26; we computed the uncertainty by combining the statistical
uncertainty from centroiding (0\farcs11) with the residual RMS scatter
(0\farcs23) in the X-ray positions of the \tycho\ reference sources
(\S\ref{sec:astrom}).  Thus, the significance of the offset is
$1.3\sigma$.  Half of this offset (0\farcs26) is attributable to the
uncertainties in centroiding and astrometry.  The remaining offset may
be due either to our underestimation of the astrometric errors or to a
complex morphology of the source on a sub-arcsecond scale
(\S\ref{sec:sgra_star_morph}).

We estimate the probability of detecting, by random chance, an
absorbed source that is as bright or brighter than \objectname[]{CXOGC
J174540.0$-$290027} and that is coincident with \sgrastar\ within
0\farcs35 as follows.  As we reported in \S\ref{sec:astrom}, we have
detected 157 sources in the 0.5--7~keV band and 71 sources in the
0.5--1.5~keV band within the ACIS-I field of view.  Selecting only
those sources that lie within a radius of 8\arcmin\ of \sgrastar\
leaves us with 143 sources in the 0.5--7~keV band and 62 sources
in the 0.5--1.5~keV band, with 24 matches between the two source
lists using a correlation radius of 2\arcsec.  After removing the
foreground sources, the resultant 0.5--7~keV source list contains 119
absorbed sources that lie within 8\arcmin\ of \sgrastar.  Of these 119
sources, only \objectname[]{CXOGC J174540.0$-$290027} was brighter
than \objectname[]{CXOGC J174540.0$-$290027} during the observation.

To determine the radial distribution of the sources on the sky, we
count up the number of sources in concentric annuli centered on
\sgrastar, using 1\arcmin-wide annuli, and fit the distribution with a
power-law model.  The best-fit radial surface density profile is given
by the equation $\sigma_{\rm X}(r) = (2.6 \pm 0.6)(r/1\arcmin)^{-1.2
\pm 0.2}$ sources per square arcminute, where $r$ is the offset angle
from \sgrastar\ in arcminutes.  Integrating the profile from 0\arcsec\
to 0\farcs35 and multiplying by 2/119, we find that the probability of
detecting, by random chance, an absorbed X-ray source that is as
bright or brighter than \objectname[]{CXOGC J174540.0$-$290027} and
that is coincident with \sgrastar\ within 0\farcs35 is $5.6 \times
10^{-3}$.  We note, however, that the radial profile given above
overpredicts the density of sources at small radii, because the source
detection efficiency of the combined HRMA/ACIS instrument drops off
with increasing off-axis angle due to the combined effects of the
increasing PSF size and the decreasing effective area.  The
integration requires an extrapolation of over an order of magnitude
toward smaller radii, so even a small flattening of the slope would
cause a significant decrease in the predicted number of sources in the
central arcsecond.  The probability given above should thus be
considered an upper limit to the true probability.  We will address
these problems in our subsequent paper on the point sources.

\subsection{Morphology}
\label{sec:sgra_star_morph}

Spatial analysis of the morphology of the source coincident with
\sgrastar\ and of two nearby point sources (\objectname[]{CXOGC
J174538.0$-$290022}\ and \objectname[]{CXOGC J174540.9$-$290014})
indicates that \sgrastar\ may be slightly extended (see
Fig.~\ref{fig:raw_central_arcmin}).  A two-dimensional Gaussian fit to
\sgrastar\ yields full-widths at half maximum (FWHM) of 1\farcs6 (E-W)
and 1\farcs3 (N-S), whereas the FWHM of the on-axis HRMA point-spread
function (PSF) is $\approx0\farcs5$.  The widths of the other two
sources are narrower (1\farcs1 and 0\farcs9 [E-W], 1\farcs0 and
0\farcs9 [N-S]), but they are still broader than the PSF by about a
factor of two.  The enlarged widths are attributed to three factors:
(1) a minor problem with the aspect solution (see below); (2) the
0.5-pixel randomization introduced into the event positions by the CXC
standard processing pipeline but not included in the PSF size
calculation, and (3) the fact that the HRMA PSF more closely resembles
a Lorentzian than a Gaussian.

Examination of the aspect solution file shows three discontinuities in
the curves recording the position of the science instrument module
(SIM) translation stage during the course of the observation.  The CXC
has determined that this problem is caused by warm pixels in the
aspect camera that sometimes fall near one of the fiducial lights used
to monitor SIM drift.  The amplitudes of the discontinuities are
$\approx 0\farcs37$ along the Z-axis of the SIM and $\approx
0\farcs15$ along the Y-axis.  The spacecraft roll angle was 268\fdg5,
so the Z- and Y-axes were aligned nearly E-W and N-S, respectively.
These discontinuities broadened the widths of all the source profiles,
but the effect on the source centroids used to determine positions on
the sky should not be a problem since all sources experienced the same
pattern.  The fact that the two comparison sources have narrower
profiles than \sgrastar\ indicates that the aspect errors cannot
account entirely for its apparent extent.  A proper study of the
spatial morphology of \sgrastar\ will require careful analysis of the
Cycle~1 data, after reprocessing to correct for the aspect problem.
We will address this in a subsequent paper in combination with our
analysis of our Cycle~2 data.

Assuming the excess extent of \sgrastar\ is real, we can obtain a
rough estimate of its size by subtracting in quadrature the mean
diameter of the two comparison sources from the mean diameter of
\sgrastar; we find that the apparent intrinsic size of \sgrastar\ is
$\approx 1\arcsec$ or 0.04~pc.  Structure in \sgrastar\ on this scale
is consistent with the expected Bondi accretion radius
\citep[1--2\arcsec][]{Bondi52} for matter accreting hydrodynamically
onto the MBH from either the stellar winds of the nearby cluster of
massive stars (\S\ref{sec:mdot}) or the hot diffuse plasma that we
observe surrounding \sgrastar\ (\S\ref{sec:hot_ism}).

\subsection{Spectroscopy}
\label{sec:spec}

\subsubsection{Continuum}
\label{sec:sgra_star_continuum}

The increased CTI in the frontside-illuminated CCDs has caused the
energy scale to become position dependent.  To correct for this, the
CXC divided each CCD into $32 \times 32$-pixel subregions and analyzed
calibration data from the on-board external calibration source to
calibrate the energy scale in each subregion.  The gain function for
each subregion is stored in a FITS embedded function (FEF) file.

We use the CIAO tool \emph{mkrmf} with an FEF
file\footnote{acis3d\_x00\_y29\_FP$-$110\_D1999-09-16fef\_phaN0002.fits}
provided by the CXC to create a spectral response matrix for analysis
of the \sgrastar\ spectrum.  An auxiliary response file describing the
energy-dependent effective area of the combined HRMA/ACIS instrument
at the location of \sgrastar\ on the I3 detector is created using the
CIAO tool \emph{mkarf} with the ACIS quantum efficiency (QE) file
provided by the CXC.\footnote{acisD1997-04-17qeN0002.fits}

A total of 258 counts are extracted in the 0.5--9~keV band from a
1\farcs5-radius circle centered on the position of the X-ray source
coincident with \sgrastar.  This aperture is small enough to minimize
contamination from several nearby sources (see
Fig.~\ref{fig:raw_central_arcmin}) yet large enough that the
percentage encircled energy from a point source at the center of the
aperture is $\ga85$\% at all energies.  A local background spectrum
with 1317 counts is extracted from a 10\arcsec-radius circle centered
on \sgrastar\ (see \S\ref{sec:diffuse_spec}); to avoid contaminating
the background spectrum with counts from the point sources in the
region, we exclude counts within a 1\farcs5 radius of \sgrastar\ or
any of the other six point sources in the extraction region (see
\S\ref{sec:pt_src_spec}).  After background subtraction, the net
counts received from \sgrastar\ in 40.3~ks are $222\pm17$ counts.  The
source is detected with high significance, $S/N \simeq 37\sigma$,
despite the highly elevated diffuse X-ray background in the central
parsec of the Galaxy (see \S\ref{sec:diffuse_spec}).

We fit the source spectrum in XSPEC with an absorbed power-law model
(see Table~\ref{tab:spec}).  The best-fit model ($\rm \chi^2/d.o.f. =
19.8/22$) has photon index $\Gamma = 2.7^{+1.3}_{-0.9}$ ($N(E) \propto
E^{-\Gamma}$ photons cm$^{-2}$ s$^{-1}$ keV$^{-1}$) and column density
$N_{\rm H} = (9.8^{+4.4}_{-3.0}) \times 10^{22}$
cm$^{-2}$.\footnote{Except where specified otherwise, the
uncertainties given in this paper are for the 90\% confidence
interval: $\Delta\chi^2 = 2.71$ for one interesting parameter.}  The
source spectrum and the best-fit absorbed power-law model are shown in
Figure~\ref{fig:sgra_star_pow_spec}.  The spectrum is binned to yield
a minimum of 10 counts per channel; this restricts the energy band of
the binned spectrum to the range 0.5 to about 7~keV.

We also fit the source spectrum with an absorbed optically thin
thermal plasma model (see Table~\ref{tab:spec}).  The optically thin
thermal plasma code that we use was developed by \citet{Raymond77}.
Twice solar abundances are assumed in the Raymond-Smith model here and
throughout this paper \citep[see][and references therein]{Morris93}.
The best-fit model ($\rm \chi^2/d.o.f. = 16.5/22$) has $kT =
1.9^{+0.9}_{-0.5}$~keV and $N_{\rm H} = (11.5^{+4.4}_{-3.1}) \times
10^{22}$ cm$^{-2}$.  The source spectrum and the best-fit absorbed
Raymond-Smith model are shown in Figure~\ref{fig:sgra_star_raym_spec}.

Both models are consistent with the data due to the low number of
counts.  Using the power-law model, the measured (absorbed) flux in
the 2--10~keV band is $(1.3^{+0.4}_{-0.2}) \times 10^{-13}$ ergs
cm$^{-2}$ s$^{-1}$, and the absorption-corrected luminosity is
$(2.4^{+3.0}_{-0.6}) \times 10^{33}$ ergs s$^{-1}$.  The thermal
plasma model gives similar numbers.  Due to the large uncertainties in
the photon index and the column density, the 2--10~keV luminosity is
known only to within a factor of two, and the extrapolated 0.5--10~keV
luminosity ($\sim 8.5 \times 10^{33}$ ergs s$^{-1}$) is uncertain by
more than an order of magnitude.  The confidence limits given for the
flux and luminosity of \sgrastar\ are derived by computing the 90\%
confidence region ($\Delta\chi^2 = 4.61$ for two interesting
parameters) for the column density versus the photon index parameters
of the absorbed power-law model with the normalization parameter of
the power law free to vary.  The column density and photon index are
then fixed at the extremum values of the 90\% confidence contour, the
spectrum is fit to determine the corresponding best-fit normalization
value, and the flux and luminosity of the model are computed.

Next, we fit the absorbed power-law model to the 0.5--7~keV spectrum
using a range of fixed column densities [$N_{\rm H} = (6, 8, 10)
\times 10^{22}$ cm$^{-2}$].  As expected the best-fit photon index
becomes flatter as the column density is decreased.  The best-fit
photon index is about 1.5 when the column density is fixed at the
canonical Galactic Center value $N_{\rm H} = 6 \times 10^{22}$
cm$^{-2}$ ($\rm \chi^2/d.o.f. = 24.2/23$).  However, a column density
this low is only marginally consistent with the data at the 99\%
confidence level.

We also fit the absorbed power-law model to the spectrum using only
the 2--7~keV and 3--7~keV ranges to minimize the effect of the column
density on the fit, but we find the resulting best-fit parameters are
highly dependent on the choice of initial parameter values; the
best-fit parameter values to the 0.5--7~keV spectrum, on the other
hand, are robust.  For the remainder of this paper, we use the
best-fit parameters derived from fits to the 0.5--7~keV spectrum (see
Table~\ref{tab:spec}).

The measured spectrum suffers from two known systematic effects.
First, the percentage encircled energy focussed by the HRMA within the
1\farcs5-radius extraction circle is energy dependent, varying from
$\approx95\%$ at 1.5~keV to $\approx85\%$ at 8.6~keV.  Second, charge
is lost as events are clocked out of detector I3 due to the increased
CTI.  This causes an energy-dependent decrease in the number of
detected events.  From measurements made with the external calibration
source at $-110\,^\circ$C, it is known that $\approx5$\% of events at
1.5~keV and $\approx20\%$ at 5.9~keV are lost.  Hence both effects
work together to lower and to steepen the spectrum systematically.
Given the numbers above, we estimate that the measured luminosity
should be increased by $\sim20$\% and the photon index should be
decreased (i.e., flattened) by $\sim0.2$--$0.3$.  These corrections
are negligible compared to the uncertainties in the model parameters
due to the small number of counts in the current data.

\subsubsection{Fe~K$\alpha$ Line}
\label{sec:sgra_star_fek}

Excess counts might be present in the 6--7~keV region of the
\sgrastar\ spectrum (see Fig.~\ref{fig:sgra_star_pow_spec}).  We test
for the presence of K$\alpha$ line emission from \ion{Fe}{25}\ by
fitting the observed spectrum with an absorbed power-law plus
gaussian-line model.  Because of the poor statistics, we fix the line
energy and width at 6.67~keV and 0.0~keV, respectively.  The best-fit
values for the remaining free parameters are $N_{\rm H} = 11.7 \times
10^{22}$ cm$^{-2}$, $\Gamma = 3.5$, $K_{\rm pl} = 1.1 \times 10^{-3}$
photons cm$^{-2}$ s$^{-1}$ keV$^{-1}$ at 1~keV, and $K_{\rm Fe} = 2.4
\times 10^{-6}$ photons cm$^{-2}$ s$^{-1}$ ($\rm \chi^{2}/d.o.f. =
15.4/21$).  Here $K_{\rm pl}$ and $K_{\rm Fe}$ are the normalization
parameters for the power-law and the Fe-line components of the model.
The best-fit equivalent width of the potential line is 1.8~keV.  For
comparison, the equivalent widths of the iron lines in the spectra of
Sgr~A East and the local diffuse emission within 10\arcsec\ of
\sgrastar\ are 3.1~keV \citep{Maeda01} and 1.3~keV (see
\S\ref{sec:diffuse_spec} and Fig.~\ref{fig:bkg_cir_spec}),
respectively.

We compute an $F$-statistic, $F_{\chi} = \Delta \chi^{2} /
\chi^{2}_{\nu2} = 6.2$, to test the significance of the additional
term in the model.  Here $\chi^{2}_{\nu2} = 0.73$ is the reduced
chi-square for the model with a line.  The probability of observing
$F_{\chi} > 6.2$ for $\nu_1 = 1$ and $\nu_2 = 21$ degrees of freedom
is 2.1\%, indicating that the significance of the improvement to the
fit from the additional term is equivalent to $2.0\sigma$ for a
Gaussian process.  Given that $\chi^{2}_{\nu2} < 1$, it is worth
examining the significance of the additional component if we set
$\chi^{2}_{\nu2} = 1$.  In that case, the probability of observing
$F_{\chi} > 4.5$ is 4.6\% or $1.7\sigma$.

An alternative possibility is that the excess counts in the \sgrastar\
spectrum result from inadequate subtraction of the iron line in the
spectrum of the local diffuse background.  However, the background
contributed only 36 of the 258 counts (14\%) in the 1\farcs5-radius
aperture used to extract the \sgrastar\ spectrum.  Furthermore, the
background-subtracted spectrum of the integrated emission from the six
point sources within 10\arcsec\ of \sgrastar\ shows no sign of excess
counts in the 6--7~keV range (see \S\ref{sec:pt_src_spec} and
Figure~\ref{fig:pt_src_spec}).  Thus, while poor counting statistics
prevent a definitive conclusion, it seems likely that an iron emission
line (or line complex) may be present in the \sgrastar\ spectrum.

\subsection{Variability}
\label{sec:sgra_star_ltc}

We examined the short-timescale temporal behavior of the \sgrastar\
X-ray source by constructing 0.5--7~keV light curves extracted from
circular regions of 1\farcs5 and 0\farcs5 radius. The larger
extraction region has 242 events, but it includes photons from an
extended region (\S\ref{sec:sgra_star_morph}) which may not vary
rapidly.  The smaller extraction region has only 67 events, but these
may arise from a compact region around the MBH.  We also examined
similarly extracted events from two unresolved sources within $\approx
0\farcm5$ of \sgrastar\ (see Fig.~\ref{fig:raw_central_arcmin}).  The
resulting light curves of \sgrastar\ and a comparison light curve of
the strongest unresolved source, \objectname[]{CXOGC
J174538.0$-$290022}, with 129 events within a 0\farcs5-radius
aperture, are shown in Figure~\ref{fig:sgra_star_ltc}.
\objectname[]{CXOGC J174538.0$-$290022} has an absorbed spectrum and
has no infrared counterpart in SIMBAD; thus it may be an accreting
X-ray binary.  The background flux from the diffuse emission
contributes 14\% and 6\% to the \sgrastar\ light curves from the
1\farcs5 and 0\farcs5 regions, respectively.

The light curve from \sgrastar\ shows a possible flare-like event
during the first hour of observation.  To evaluate the statistical
significance of the event, we calculate the nonparametric
Kolmogorov-Smirnov statistic to test the hypothesis that the source is
constant.  The probability of constancy for the \sgrastar\ X-ray flux
is $P = 8.8 \times 10^{-2}$ for the 1\farcs5 extraction circle and $P
= 7.6 \times 10^{-3}$ for the 0\farcs5 extraction circle.  The latter
value is equivalent to a $2.7\sigma$ event for a Gaussian process.
For comparison, the probability of constancy for the control source
\objectname[]{CXOGC J174538.0$-$290022} in the 0\farcs5 extraction
circle is as high as $P = 7.3 \times 10^{-1}$, indicating no
significant variability.  The other control source \objectname[]{CXOGC
J174540.9$-$290014} has a probability of constancy in the 0\farcs5
extraction circle of $6.0 \times 10^{-2}$, which is again consistent
with no significant variability.  Only \sgrastar\ shows the
possibility of variability above the equivalent $2\sigma$ level in the
0\farcs5 extraction circle.

We thus find highly suggestive, but inconclusive, evidence of rapid
X-ray variability from a compact component within the \sgrastar\
source.  The peak luminosity of the putative flare is $L_{\rm X}
\simeq 6 \times 10^{33}$ ergs s$^{-1}$.  Additional data are needed to
establish whether or not the variability is real.  The search for
rapid X-ray variability of \sgrastar\ is of crucial importance, since
it has the potential to provide a powerful discriminator between MBH
and stellar origins for the X-ray source and between the various
proposed MBH accretion-flow emission processes for \sgrastar.

\section{Integrated Point Source Emission in the Central Parsec}
\label{sec:pt_src_spec}

Figure~\ref{fig:pt_src_spec}\ shows the spectrum of the integrated
X-ray emission from the 6 point sources observed within 10\arcsec\ of
\sgrastar\ (see \S\ref{sec:sgra_star_continuum} and
Fig.~\ref{fig:raw_central_arcmin}).  The solid line in the upper panel
of Figure~\ref{fig:pt_src_spec}\ is the best-fit absorbed power-law
model ($\rm \chi^2/d.o.f. = 56.3/72$).  The parameters of the model
are listed in Table~\ref{tab:spec}, along with the integrated flux and
luminosity.  The net count rate from the six sources is $(1.4 \pm 0.1)
\times 10^{-2}$ counts s$^{-1}$ (0.5--7~keV).  These sources are
discussed further in \S\ref{sec:confusion}.

The spectrum shows no obvious signs of emission lines.  To quantify
this statement, we fit the spectrum with an absorbed Raymond-Smith
model.  The best-fit column density is implausibly high ($N_{\rm H} =
(24.1^{+5.0}_{-4.4}) \times 10^{22}$ cm$^{-2}$) when the elemental
abundances are fixed at twice the solar abundances ($\rm
\chi^2/d.o.f. = 70.9/72$).  When the metallicity is allowed to vary,
the best-fit column density is reasonable ($N_{\rm H} =
(12.3^{+4.4}_{-2.7}) \times 10^{22}$ cm$^{-2}$), but the abundances
are set to zero by the fitting engine ($\rm \chi^2/d.o.f. = 56.4/71$).

\section{Diffuse X-ray Emission in the Central Parsec}
\label{sec:diffuse_spec}

As discussed in \S\ref{sec:gcimage}, the entire Sgr~A complex ---
comprised of Sgr~A East, Sgr~A West, and \sgrastar --- sits on a ridge
of X-ray emission extending north and east parallel to the Galactic
plane (see Fig.~\ref{fig:gal_center}).  To study the spectrum of the
diffuse emission within 10\arcsec\ of \sgrastar\ (hereafter the local
diffuse emission), it is necessary to first subtract off this
underlying background.  The structure of the emission along the
Galactic plane is complex, making it difficult to determine a proper
estimate of the background near \sgrastar.  We selected a region about
42\arcsec\ north of \sgrastar\ that lies outside the intense X-ray
emission from the Sgr~A complex and yet within the extended radio
structure known as the Sgr~A ``halo'', where the X-ray surface
brightness is relatively flat.  We extracted a spectrum with 544 total
counts from a circular region 15\arcsec\ in radius centered at $\rm
17^h45^m39\fs7$, $-28\arcdeg59\arcmin47\arcsec$ (J2000.0).  This
extraction circle lies entirely on the I3 detector, as does the
extraction circle for the local diffuse emission.  The total count
rate in the 0.5--7~keV band from this region of the Galactic plane
emission is $(1.9 \pm 0.1) \times 10^{-5}$ counts s$^{-1}$
arcsec$^{-2}$.

Next we analyzed the spectrum of the local diffuse emission, using the
Galactic plane spectrum for the background.
Figure~\ref{fig:bkg_cir_spec}\ shows the background-subtracted
spectrum.  An emission line from highly ionized iron is clearly
visible at 6.7~keV, indicating that most of the emission comes from a
hot optically thin thermal plasma.  We therefore fit the spectrum with
an absorbed Raymond-Smith model.  The best-fit model is indicated by
the solid line in the upper panel of Figure~\ref{fig:bkg_cir_spec},
and the best-fit parameters are listed in Table~\ref{tab:spec}.  The
2--10~keV flux and luminosity of the local diffuse emission are $(1.9
\pm 0.1) \times 10^{-15}$ ergs cm$^{-2}$ s$^{-1}$ arcsec$^{-2}$ and
$(7.6^{+2.6}_{-1.9}) \times 10^{31}$ ergs s$^{-1}$ arcsec$^{-2}$; the
net count rate is $(1.2 \pm 0.1) \times 10^{-4}$ counts s$^{-1}$
arcsec$^{-2}$ (0.5--7~keV).  The equivalent width of the iron line is
$\sim1.3$~keV.

Based on the parameters of the best-fit model, we estimate that the
local, hot diffuse plasma has an RMS electron density $\langle n_{\rm
e}^2 \rangle^{1/2} \approx 26$ cm$^{-3}$ and emission measure $EM
\approx 540$ cm$^{-6}$ pc.  Here we have assumed the plasma has unity
filling factor and is fully ionized with twice solar abundances (mean
atomic weight $\mu = 0.70$).  The total mass of this gas is $M_{\rm
local} \approx 0.1$~\msun.

The local plasma around \sgrastar\ appears to be somewhat cooler ($kT
= 1.3^{+0.2}_{-0.1}$~keV) than the plasma we analyzed in our companion
study of Sgr~A East \citep{Maeda01}.  Using the MEKA thermal plasma
model developed by \citet{Mewe85}\ and \citet{Kaastra92},
\citeauthor{Maeda01}\ find the Sgr~A East plasma to have $kT =
2.1^{+0.3}_{-0.2}$~keV.  The column densities derived from the fits to
both plasmas are consistent at $\sim 1.2 \times 10^{23}$ cm$^{-2}$.
We allowed the elemental abundances to vary when fitting the high
signal-to-noise spectrum of Sgr~A East.  The best-fit model indicates
that the Sgr~A East plasma has about 4 times solar abundances.
Fitting the local emission near \sgrastar\ with the abundances fixed
at four times the solar value did not significantly change the
best-fit temperature.  The difference in temperature between the two
plasmas therefore appears real.

In addition, \citeauthor{Maeda01}\ find that the net count rate within
a 40\arcsec-radius circle centered on Sgr~A East at the position $\rm
17^h45^m44\fs1$, $-29\arcdeg00\arcmin23\arcsec$ (J2000.0) is $\approx
3.6 \times 10^{-5}$ counts s$^{-1}$ arcsec$^{-2}$.  Thus, the net
count rate from the local diffuse emission is 3.3 times the net count
rate within Sgr~A East and 6.3 times the total count rate in the
background region along the Galactic plane.  This shows that the
diffuse X-ray emission from the Sgr~A complex is significantly peaked
around \sgrastar.

It is likely that some fraction of the local diffuse emission is
contributed by the stars in the central parsec cluster.
\citet{Genzel96}\ estimate that the core radius of the cluster is
$\sim 0.4$~pc (10\arcsec) with a stellar mass density in the core of
$\sim 4 \times 10^{6}$ \msun\ pc$^{-3}$.  The 2--10~keV luminosity of
$2.4 \times 10^{34}$ ergs s$^{-1}$ could arise from the ordinary OB
and Wolf-Rayet (WR) stars present in the cluster.  This level of
emission is an order of magnitude higher than seen in young stellar
clusters like the Orion Trapezium or W3 but is comparable to that seen
from the R136a cluster and associated WR stars in 30~Doradus
\citep{Feigelson01}.

Throughout this paper we have neglected the instrumental background in
our analysis for two reasons.  First, \citet{Koyama89}\ observed the
center of our Galaxy with \ginga\ and found K$\alpha$ transition lines
from highly ionized ions of iron extending over the central 100~pc.
They also found diffuse continuum emission on this scale.  They
interpret both emission features as coming from an optically thin
$\sim10$~keV plasma.  This diffuse X-ray emission extends far beyond
the ACIS-I field of view; thus we cannot measure the instrumental
background spectrum directly from our data.  Second, the expected
count rate from the instrumental background is negligible compared to
the contributions from the bright diffuse emission in the field, as we
now show.

The non-X-ray background rate measured by the CXC in \chandra/ACIS
observations of high Galactic-latitude fields is $\sim8 \times
10^{-7}$ counts s$^{-1}$ arcsec$^{-2}$ in the 0.5--7~keV band (see
\S6.10 of the POG).  For comparison, the average count rate in ACIS
detector S2 within a 2\arcmin-radius circle centered at $\rm
17^h46^m46\fs8$, $-29\arcdeg05\arcmin57\arcsec$ (J2000.0) is $(1.7 \pm
0.8) \times 10^{-6}$ counts s$^{-1}$ arcsec$^{-2}$.  This region is
devoid of strong point sources in the data and is located 15\arcmin\
off axis, where the effective area at 2.4~keV is only about 75\% of
the on-axis value; it thus provides us with a useful upper-limit to
the instrumental background rate in the four ACIS-I CCDs during the
observation.  The expected non-X-ray background rate is seen to be
about half the observed background rate in detector S2.  The excess
count rate may be attributed to the diffuse X-ray emission from hot
gas along the line of sight through the center of our Galaxy.
Referring back to the count rate from the local diffuse emission given
above, it can be seen that the instrumental background contributes
only $\sim0.7\%$ of the background counts in the vicinity of
\sgrastar; it is therefore negligible for our purposes in this paper.

\section{Stellar Counterparts to Detected X-ray Sources and Source
Confusion in the Central Parsec}
\label{sec:confusion}

In addition to detecting X-rays from a source coincident with
\sgrastar\ within 16 light days for the first time, we have resolved
the diffuse X-ray emission from Sgr~A East and along the Galactic
plane and detected over 150 point sources in the $17\arcmin \times
17\arcmin$ ACIS-I field of view.  For comparison, the \rosat/PSPC
detected 14 sources in a $30\arcmin \times 30\arcmin$ field around the
Galactic Center in an observation of equal duration \citep{Predehl94}.
The \einstein/IPC, with a harder energy band than \rosat, but less
effective area, detected 12 sources in about 20~ks in a $\sim1\arcdeg
\times 1\arcdeg$ field \citep{Watson81}.

\begin{figure}[ht]
%\epsscale{1.0}
\plotone{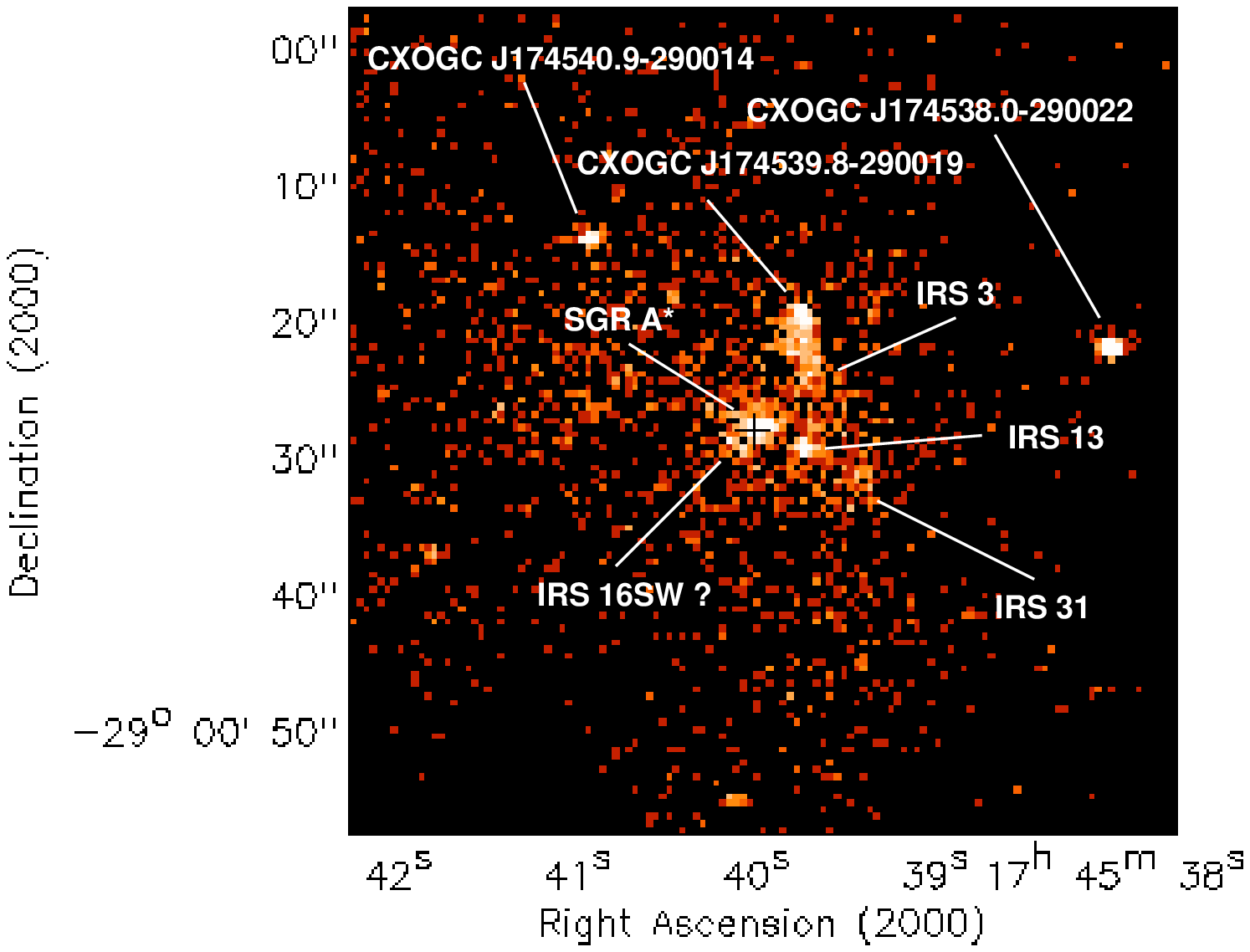}
\caption{\chandra\ 0.5--7~keV image of the central $1\arcmin \times
1\arcmin$ of the Galaxy (\S\ref{sec:sgra_star_position}).  This image
has not been smoothed or flat-fielded.  Each pixel subtends a solid
angle of $0\farcs5 \times 0\farcs5$ on the sky.  The black cross marks
the radio position of \sgrastar\ as determined by
\citet{Yusef-Zadeh99}.  The cross lies superposed on the X-ray source
that we associate with \sgrastar\ based on the extremely close
positional coincidence.  Tentative identifications with bright IR
sources are shown for several of the point sources that were detected
using the wavelet source detection algorithm developed by the Chandra
X-ray Center (\S\ref{sec:astrom}).  The question mark next to the
label IRS~16SW indicates the possible detection of an X-ray source
that would coincide with IRS~16SW within 1--2\arcsec.  There appears
to be a significant excess of counts at this location, but the wavelet
algorithm did not identify a source there, possibly due to its
proximity to the brighter source at the position of \sgrastar.  No
matches were found in the SIMBAD database within a 3\arcsec\ radius
for two of the brightest sources in this field: CXOGC
J174540.9$-$290014 and CXOGC J174538.0$-$290022.  These sources are
likely candidates for new X-ray binaries.  Bright diffuse emission
from hot gas is visible throughout the region.  It is possible that
some of this emission may come from a collection of faint point
sources.
\label{fig:raw_central_arcmin}}
\end{figure}

\begin{figure}[ht]
%\epsscale{1.0}
\plotone{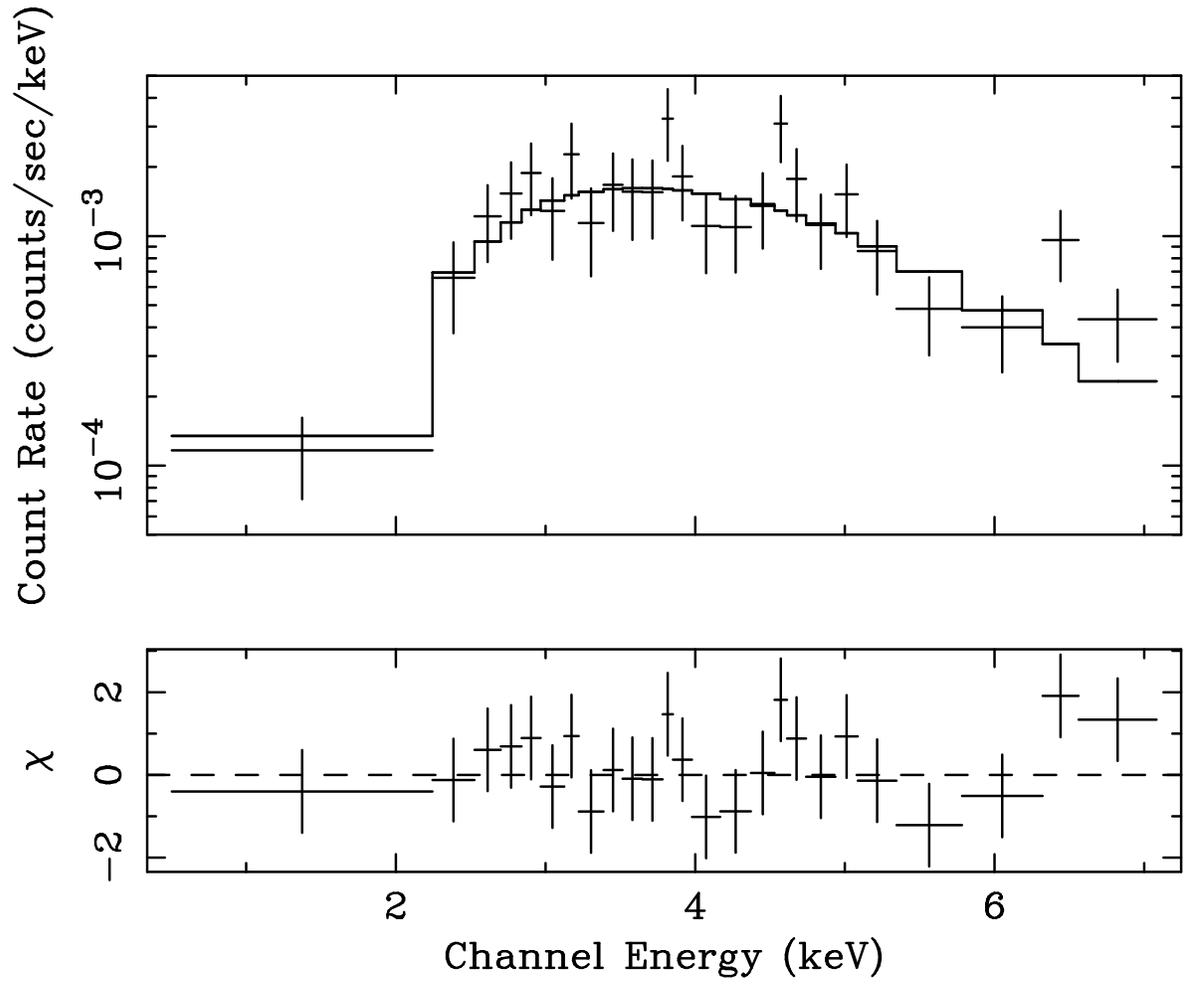}
\caption{X-ray spectrum of \sgrastar\ with the best-fit absorbed
power-law model (solid line in upper panel,
\S\ref{sec:sgra_star_continuum}).  The residuals of the fit are shown
in the lower panel.  The parameters of the best-fit model are
presented in Table~\ref{tab:spec}.
\label{fig:sgra_star_pow_spec}}
\end{figure}

\begin{figure}[ht]
%\epsscale{1.0}
\plotone{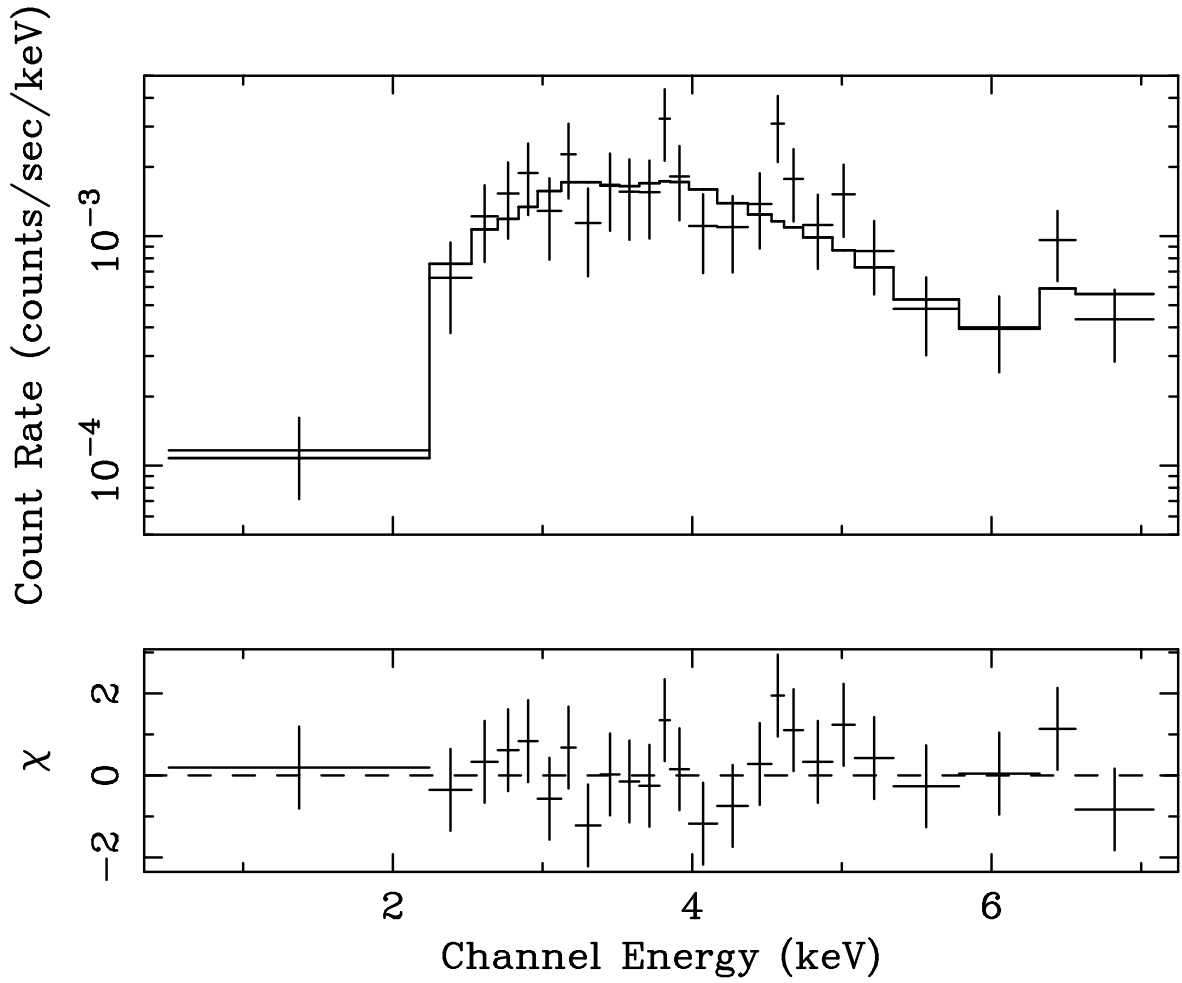}
\caption{X-ray spectrum of \sgrastar\ with the best-fit absorbed
Raymond-Smith thermal plasma model (solid line in upper panel,
\S\ref{sec:sgra_star_continuum}).  The residuals of the fit are shown
in the lower panel.  The parameters of the best-fit model are
presented in Table~\ref{tab:spec}.
\label{fig:sgra_star_raym_spec}}
\end{figure}

\begin{figure}[ht]
%\epsscale{1.0}
\plotone{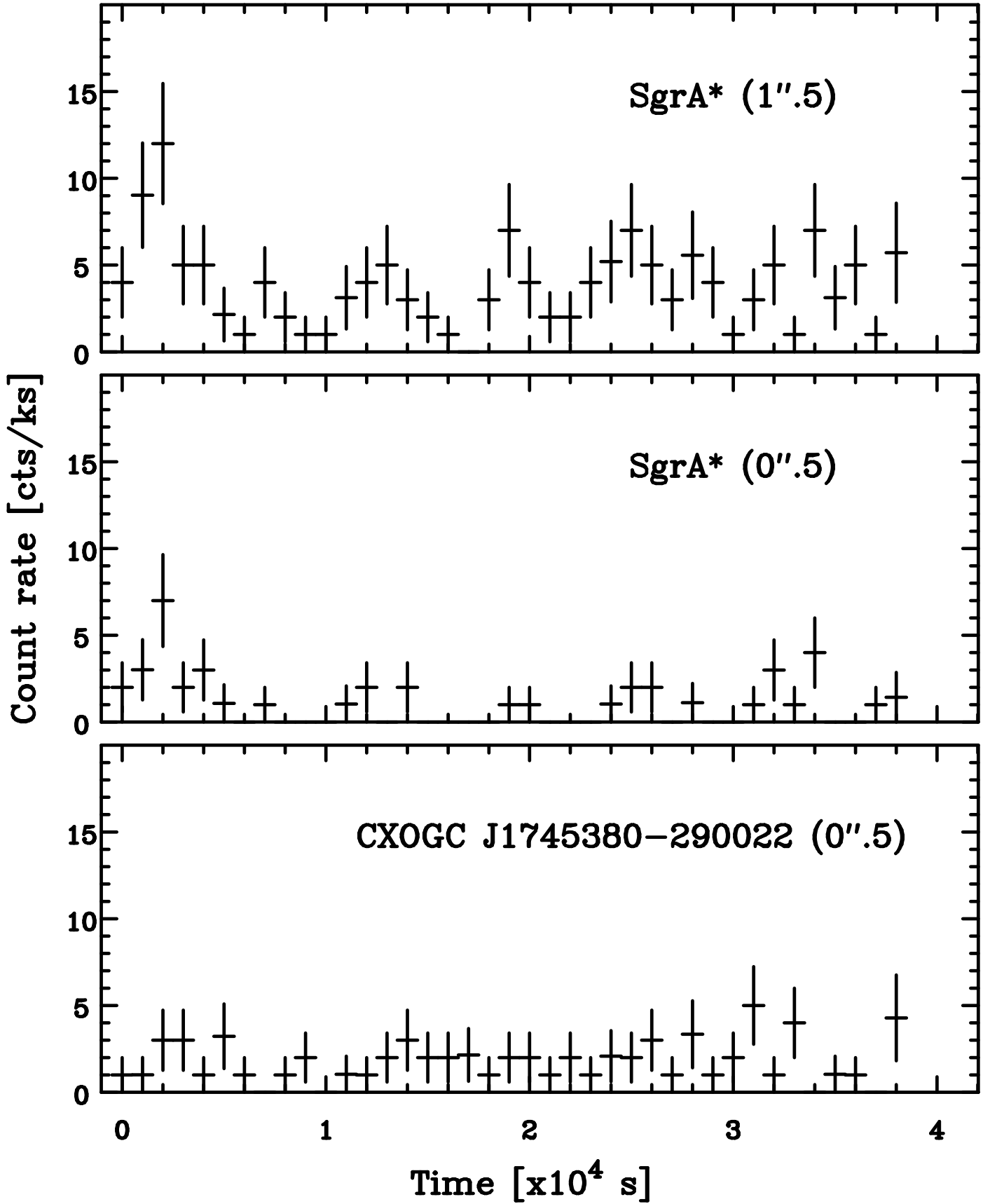}
\caption{\chandra\ 0.5--7~keV light curves of \sgrastar\ and a nearby
unresolved comparison source with 1~ks bins. {\it Top:} \sgrastar\
events from a circular region with 1\farcs5 radius. {\it Middle:}
\sgrastar\ events from a 0\farcs5-radius region.  {\it Bottom:} CXOGC
J174538.0$-$290022 events from a 0\farcs5-radius region.  Note the
possible ``flare-like'' feature at the beginning of the \sgrastar\
light curve (see \S\ref{sec:sgra_star_ltc}\ for details).
\label{fig:sgra_star_ltc}}
\end{figure}

\begin{figure}[ht]
%\epsscale{1.0}
\plotone{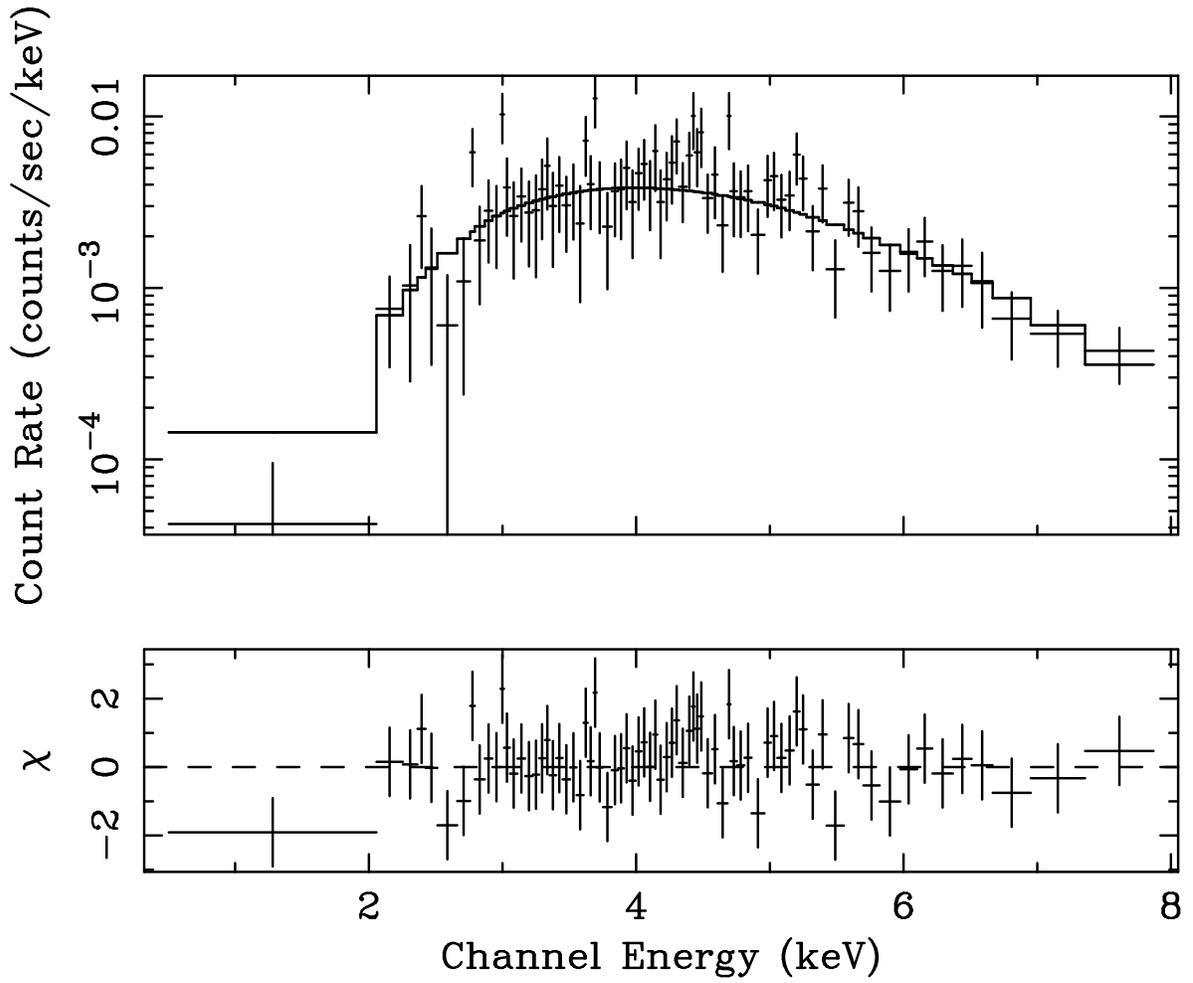}
\caption{X-ray spectrum of the integrated emission from six point
sources within a 10\arcsec\ radius of \sgrastar\
(\S\ref{sec:pt_src_spec}).  The solid line in the upper panel is the
best-fit absorbed power-law model.  The residuals of the fit are shown
in the lower panel.  The parameters of the best-fit model are
presented in Table~\ref{tab:spec}.  Note the absence of any feature
from Fe-line emission in the 6--7~keV range. \label{fig:pt_src_spec}}
\end{figure}

\begin{figure}[ht]
%\epsscale{1.0}
\plotone{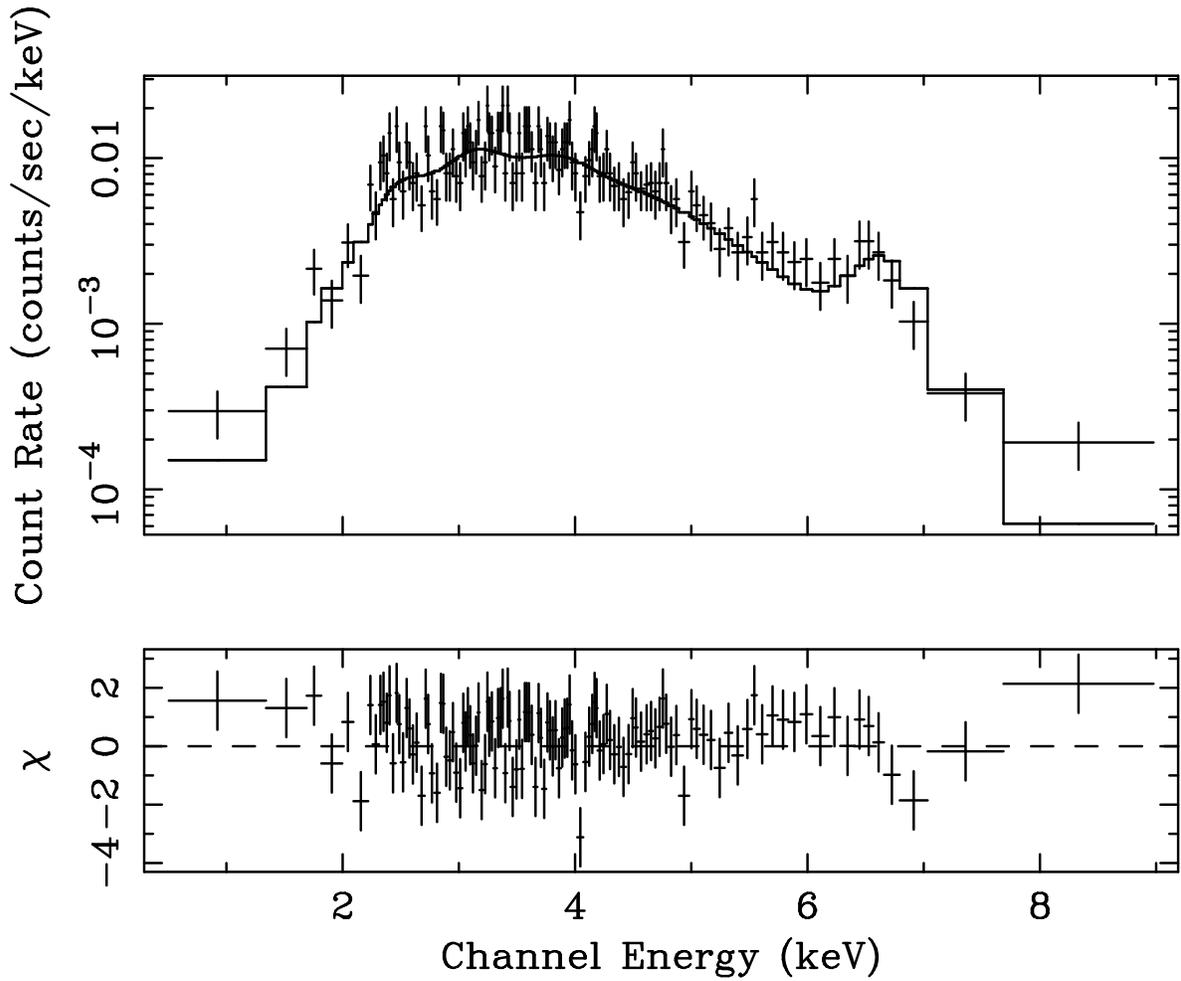}
\caption{X-ray spectrum of the local diffuse background emission
within a 10\arcsec\ radius of \sgrastar\ (\S\ref{sec:diffuse_spec}).
The solid line in the upper panel is the best-fit absorbed
Raymond-Smith thermal plasma model.  The residuals of the fit are
shown in the lower panel.  The parameters of the best-fit model are
presented in Table~\ref{tab:spec}.  Note the presence of strong
Fe-line emission in the 6--7~keV range.
\label{fig:bkg_cir_spec}}
\end{figure}

As discussed in \S\ref{sec:pt_src_spec}\ and \S\ref{sec:diffuse_spec},
several point-like X-ray sources lie within 10\arcsec\ of \sgrastar\
(see Fig.~\ref{fig:raw_central_arcmin} and Table~\ref{tab:spec}), and
the diffuse emission is also quite prominent ($F_{\rm X}$[2--10~keV]$
\approx 2 \times 10^{-15}$ ergs cm$^{-2}$ s$^{-1}$ arcsec$^{-2}$).
Our ACIS observation shows that on 1999 September 21, \sgrastar\
contributed only 12\% of the 2--10~keV flux within this region of the
sky.  All of the emission from this region would have been unresolved
by the \rosat/PSPC, which had a spatial resolution of 10--20\arcsec\
(FWHM).  Most of it will fall within the \xmm\ beam [6\arcsec\ (FWHM),
15\arcsec\ (HPD)].

The source to the southwest of \sgrastar\ matches the radio position
of \objectname[]{IRS~13} to within $\approx 1\arcsec$.
\objectname[]{IRS~13} is known to consist of a complex of stars and a
diffuse source from a strong shock at the edge of the ``mini-cavity''
seen in radio and mid-IR images.  \citet{Paumard01}\ have taken a
high-resolution IR spectrum at 2.06~\micron\ which shows that
\objectname[]{IRS~13E3} is a \ion{He}{1}\ emission-line star with a
broad P~Cygni line profile.  They propose that \objectname[]{IRS~13E3}
is one of a group of stars in the central parsec cluster that are in
the WR stage.  There is also a hint of excess counts from the vicinity
of \objectname[]{IRS~16SW} (see Fig.~\ref{fig:raw_central_arcmin}),
although no source was found there by \emph{wavdetect}, perhaps due to
its faintness and proximity to the brighter X-ray source located at
the position of \sgrastar.  \citet{Ott99}\ claim that
\objectname[]{IRS~16SW} is probably an eclipsing He-star binary,
raising the possibility that we may be seeing X-rays from their
colliding stellar winds.  Emission at levels of order $10^{32-33}$
ergs s$^{-1}$ in the 2--10~keV band is well established in WR stars
(e.g., HD~193793, \citealt{Koyama90,Koyama94}; V444~Cyg,
\citealt{Maeda99}), where the hard component is attributed to
colliding winds in a close binary system.

We have not detected other members of the \objectname[]{IRS~16}
cluster, which is known to contain a number of \ion{He}{1}\
emission-line stars.  \citet{Paumard01}\ find that
\objectname[]{IRS~16NE}, \objectname[]{IRS~16C},
\objectname[]{IRS~16SW}, and \objectname[]{IRS~16NW} are He stars with
narrow P~Cygni line profiles and propose that they are in or near the
luminous blue variable (LBV) phase.  Such stars have substantially
weaker hard X-ray emission than colliding-wind WR binaries.

The apparently diffuse emission located about 7\arcsec\ northwest of
\sgrastar\ in Figure~\ref{fig:raw_central_arcmin}\ does not correspond
with any excess of radio emission in a VLA 6-cm map of the region made
by F.~Yusef-Zadeh (1999, private communication, see
Fig.~\ref{fig:sgra_west}); on the contrary, there seems to be an
absence of radio emission at this location in the radio map.  The same
is true in the mid-IR (M.~Morris, in preparation).  We attribute this
structure to emission from a group of 3 or more point sources located
along a line running approximately north-south and covering a distance
of about 7\arcsec.  The brightest source stands out in
Figure~\ref{fig:sgra_west}\ as the red dot at the northern end of the
structure.

The stellar identifications shown in
Figure~\ref{fig:raw_central_arcmin} are \emph{tentative}; they are
based \emph{solely} on positional coincidence at this time.  The X-ray
sources marked \objectname[]{IRS~3} and \objectname[]{IRS~13} match
the coordinates listed in SIMBAD to within 1\arcsec, while
\objectname[]{IRS~31} coincides only within 2\arcsec, so this latter
match is not compelling.  The multiple IR source \objectname[]{IRS~15}
is the nearest known source to the X-ray source \objectname[]{CXOGC
J174539.8$-$290019}, though it is displaced by 2--3\arcsec, so it is
an unlikely counterpart \citep[IRS~15SW is a He star with a broad-line
profile,][]{Paumard01}.  Interestingly, no matching IR source was
found within 3\arcsec\ of \objectname[]{CXOGC J174538.0$-$290022},
despite the fact that it is the second brightest X-ray source in the
entire field and the brightest absorbed source.  Likewise, no match
was found within 3\arcsec\ of \objectname[]{CXOGC J174540.9$-$290014}.
These two sources are therefore likely candidates for X-ray binaries.

\section{A Limit on the Long-term X-ray Variability of \sgrastar}
\label{sec:long_term_var}

The center of our Galaxy has been observed by a series of X-ray
satellites over the past twenty years (\S\ref{sec:x-ray}).  Prior to
\chandra, the highest angular resolution observations were made by the
PSPC and the HRI instruments on \rosat\ \citep{Predehl94,Predehl96}.
The HRI did not detect a source at the position of \sgrastar\ in a
27~ks observation.  The PSPC detected a source, RX~J1745.6$-$2900, in
March 1992 that was coincident with \sgrastar\ within 10\arcsec.
The 0.8--2.5~keV luminosity of the source was $\sim7 \times 10^{35}$
ergs s$^{-1}$.  Hard X-ray (2--30~keV) observations made with
non-imaging instruments in the late 1980's and early 1990's,
especially the ART-P telescope on \granat, showed a long-term variable
source in the vicinity of \sgrastar\ with 2--10~keV luminosity ranging
from (2--$10) \times 10^{35}$ ergs s$^{-1}$ \citep[but see Maeda et
al.\ 1996 for an alternative
explanation]{Skinner87,Kawai88,Pavlinsky94}.  \sax\ observed the Sgr~A
complex in August 1997 and placed an upper limit on the 2--10~keV
luminosity of \sgrastar\ of $\la 10^{35}$ ergs s$^{-1}$
\citep{Sidoli99}.  Taken together, these observations suggest that
\sgrastar\ might be a variable X-ray source (by a factor of 10 or
more) and that it might have been as luminous as a few $\times
10^{36}$ ergs s$^{-1}$ within the past 15 years.  Alternatively, given
the source confusion described in \S\ref{sec:confusion}, it is
possible that \sgrastar\ was not detected by these X-ray satellites
because it was too faint.

It would be difficult to compare properly the count rate from
\sgrastar\ measured with \chandra/ACIS-I in September 1999 with the
count rates measured by the previous instruments with angular
resolutions of order 1\arcmin\ or larger, since this would require
estimating the count rates of many potentially variable point sources
in the field.  The relatively high spatial resolution of the
\rosat/PSPC, on the other hand, makes such a comparison reasonably
straightforward.  Furthermore, the \rosat\ observation occurred within
half a year of the fourth of a series of semi-annual ART-P
observations in 1990--91 that detected a source in the vicinity of
\sgrastar\ with a persistent hard X-ray luminosity of $\approx10^{36}$
ergs s$^{-1}$ and variability by about a factor of two on a half-year
timescale \citep[see Table~1 in][]{Predehl94}.  To simultaneously
reproduce the low-energy spectrum measured with the PSPC and the
higher energy spectrum measured with the ART-P,
\citeauthor{Predehl94}\ find the absorption column to the source has
to be $N_{\rm H} \approx (1.5$--$2) \times 10^{23}$ cm$^{-2}$.  This
assumed column density is consistent, within the uncertainties, with
the column density measured by ACIS.  The PSPC observation may thus
allow an indirect comparison to the ART-P observations as well.

Adopting an absorption column of $1 \times 10^{23}$ cm$^{-2}$, we
re-fit in XSPEC the spectra of the three emission components listed in
Table~\ref{tab:spec} and used the best-fit models with the response
matrix \emph{pspcb\_gain2\_256.rsp} to compute the predicted PSPC
count rate for each component in the 0.8--2.5~keV band.  To
convert the surface brightness of the local diffuse emission into an
expected count rate, we assumed the source counts were extracted from
a circular region of radius 20\arcsec; for comparison, the 50\%
encircled energy radius of the PSPC was about 15--20\arcsec.  The
predicted PSPC count rates are $2 \times 10^{-5}$ counts s$^{-1}$ for
\sgrastar, $3 \times 10^{-5}$ counts s$^{-1}$ for the summed point
sources, and $6.9 \times 10^{-4}$ counts s$^{-1}$ for the local
diffuse emission.  Summing these contributions, we find the total
predicted PSPC count rate in September 1999 would be $7.4 \times
10^{-4}$ counts s$^{-1}$.  The actual PSPC count rate observed in
March 1992 was $8 \times 10^{-4}$ counts s$^{-1}$
\citep{Predehl94}, consistent with the ACIS-based prediction.

Assuming the flux of the point sources and the local diffuse emission
remained constant between the two epochs, we find that the $3\sigma$
upper-limit on the count rate of \sgrastar\ in March 1992 is $4.6
\times 10^{-4}$ counts s$^{-1}$.  Taking into account the factor of
two uncertainty in the 2--10~keV luminosity of \sgrastar\ measured by
\chandra, the corresponding upper-limit on the luminosity of
\sgrastar\ in 1992 is $L_{\rm X} \sim (1$--$2) \times 10^{35}$ ergs
s$^{-1}$.  The 2--10~keV luminosity measured with ART-P in autumn of
1991 was $\sim8 \times 10^{35}$ ergs s$^{-1}$, which is still a factor
of 4--8 times higher than the upper limit in 1992.  Based on this
analysis and the factor of two variability seen by ART-P over a 2-year
period, it would appear that the PSPC should have seen at least an
order of magnitude higher count rate if \sgrastar\ were as luminous as
$\sim10^{36}$ ergs s$^{-1}$ in the ART-P energy band in late 1991.
This suggests that one of the other point sources within the ART-P
beam may have been responsible for the observed variation in
luminosity.  A likely candidate for the contaminating source is
AX~J1745.6$-$2901, a low-mass X-ray binary discovered with \asca\ by
\citet{Maeda96}.

While we cannot exclude the possibility that \sgrastar\ was as
luminous as $L_{\rm X} \sim 10^{35}$ ergs s$^{-1}$ during the past 15
years, all previous data are consistent with the much lower luminosity
of $L_{\rm X} \sim 2 \times 10^{33}$ ergs s$^{-1}$ that we measured in
1999 with \chandra/ACIS-I.

\section{Origin of the X-rays Coincident with \sgrastar: MBH or Stellar?}
\label{sec:mbh_or_stellar}

The MBH at \sgrastar\ is embedded in a rich and massive cluster of
very luminous stars.  Sixteen early-type, \ion{He}{1}/\ion{H}{1}\
emission-line stars with strong winds have been spectroscopically
identified within a radius of $\approx 10\arcsec$ around \sgrastar\
\citep{Krabbe95,Najarro97,Paumard01}.  Such stars are thought to be
close cousins to stars in the LBV phase and the WR stage, although
their nature is not completely determined.  As WR stars are
significant X-ray emitters, especially those in close binaries with
other WR or O stars, one must consider whether the emission we see at
\sgrastar\ arises from MBH or stellar processes.  For instance,
\citet{Ozernoy97} predict that variable X-ray emission with $L_{\rm X}
\sim 10^{33-35}$ ergs s$^{-1}$ should be present in the Galactic
Center cluster due to colliding stellar winds.

We consider the X-ray properties described in \S\ref{sec:sgra_star}.
The position of \objectname[]{CXOGC J174540.0$-$290027} is consistent
with the radio position of \sgrastar\ at the $1.3\sigma$ level, where
the $1\sigma$ positional uncertainty is 0\farcs26.  In addition, the
source has an apparent diameter of $\approx 1\arcsec$.  The stellar
cluster is a composite structure with a dense compact component
5\arcsec\ in diameter lying within a larger 20\arcsec-diameter
component \citep{Eckart99,Paumard01}, but the compact component (the
\objectname[]{IRS~16} complex) is centered 2\arcsec\ east of
\sgrastar\ and of our X-ray source.

The \ion{He}{1}\ emission-line stars nearest to \sgrastar\ (in
projection) are \objectname[]{IRS~16C}, \objectname[]{IRS~16NW}, and
\objectname[]{IRS~16SW}.  As noted in \S\ref{sec:confusion}, an excess
of counts appears around the position of \objectname[]{IRS~16SW}, but
no X-ray sources are visible in the current data at the positions of
\objectname[]{IRS~16C} and \objectname[]{IRS~16NW}.  Importantly, no
bright \ion{He}{1}\ emission-line star lies closer than 1\farcs2 to
\sgrastar\ \citep{Krabbe95,Paumard01}, which is strong evidence
against \objectname[]{CXOGC J174540.0$-$290027} being emission from a
\ion{He}{1}\ star.

The X-ray emission from single OB and WR stars is typically quite
soft, with $kT < 1$~keV, and consequently cannot be observed at the
Galactic Center due to the obscuration.  Close binary WR+WR and WR+O
systems, in which the X-rays arise from colliding winds, can show
harder spectra with $kT \simeq 1$--3~keV and $L_{\rm X} \simeq
10^{32-33}$ ergs s$^{-1}$ in the 2--10~keV band
\citep[e.g.,][]{Corcoran96,Maeda99}.  The spectral and luminosity
characteristics for the more extreme WR binaries are roughly
consistent with those of the source coincident with \sgrastar.

The variability tentatively reported in \S\ref{sec:sgra_star_ltc}, if
confirmed, is not consistent with the behavior of WR binaries.
Variations associated with binary phase are typically seen on
timescales of days to years \cite[e.g.,][]{Williams90}.  Variations on
timescales of $\sim 1$ hour with amplitudes of $L_{\rm X} \approx
10^{33}$ ergs s$^{-1}$, as might have been seen in \sgrastar, are
unprecedented among colliding wind binaries.  A firm detection of such
rapid variability would provide additional strong evidence against a
WR star origin for the emission from \objectname[]{CXOGC
J174540.0$-$290027}.

The colliding winds model for X-ray emission \citep{Ozernoy97}
requires that, in order to reproduce the X-ray luminosity of
\objectname[]{CXOGC J174540.0$-$290027}, the stars must be much closer
to each other (a few $\times 10^{14}$ cm) than the typical separation
of the \ion{He}{1}\ emission line stars in the central parsec (a few
$\times 10^{17}$ cm).  However, Ozernoy et al.\ raise the possibility
that a sizeable population of OB stars may be present in the cluster,
and that the X-rays arise in the shocks produced at the interfaces of
the winds of these OB stars and the WR-type emission-line stars.  The
number of O stars required for a substantial probability of a
sufficiently close encounter is $\ga 10^6$, however, far larger than
the luminosity constraints allow, so one must appeal to WR+OB binary
systems for anything but an occasional X-ray flare of several weeks
duration.  The most significant constraint on the WR+OB colliding wind
model is that there is no known WR star coincident with
\objectname[]{CXOGC J174540.0$-$290027}.

In addition to the central cluster of emission-line stars, the central
few hundredths of a parsec surrounding \sgrastar\ ($\approx 0\farcs5
\times 0\farcs5$) contains a concentration, or cusp, of at least a
dozen bright stars (K $\simeq$ 14--16 mag), which, according to
\citet{Eckart99}, are predominantly blue and featureless, indicating
that they may be O stars.  This ``\sgrastar\ (IR)'' cluster warrants
consideration as the source of the X-rays observed toward \sgrastar\
if the winds from these stars are typical of those of O stars, because
colliding O-star winds can also generate measurable X-ray fluxes
\citep[c.f.,][]{Pittard97}, and the size of this cusp of stars can
roughly account for the observed extent of \objectname[]{CXOGC
J174540.0$-$290027}.

The typical separation of the stars observed in the central cusp is
about 0\farcs1, or $\sim 10^{16}$ cm.  The calculations of
\citeauthor{Pittard97} indicate that a separation $\la 10^{14}$ cm is
needed, even in the most favorable case, to reproduce the luminosity
observed for \objectname[]{CXOGC J174540.0$-$290027}.  Therefore, one
must again invoke close binary systems or expect only rare and brief
events.  Nothing is currently known concerning the binarity of the
stars in the cusp; in this dense stellar environment, the dynamical
evolution of binary systems should be relatively rapid.  We note that
the observed spectra of the O+O wind binaries HD~57060 and $\delta$
Orionis that \citeauthor{Pittard97} compare to their models have $kT <
1$~keV, which would be unobservable at the Galactic Center.  The
possibility that some or all of the flux of \objectname[]{CXOGC
J174540.0$-$290027} is due to colliding winds remains open and can be
investigated using the source variability and spectrum.

One can also consider an origin from young lower-mass stars, which are
likely to be present among the luminous young OB/WR stars in the
stellar cluster.  Late-type stars have X-ray emission elevated by
factors of $10^{1-4}$ above their main sequence levels during their
first $10^7$ years due to enhanced magnetic activity
\citep{Feigelson99}.  In two observed cases, X-ray flares exhibited
peak $L_{\rm X} \simeq 1$--$2 \times 10^{33}$ ergs s$^{-1}$, with $kT
\simeq 7$--10~keV, and decays on timescales of hours
\citep{Preibisch95, Tsuboi98}.  It is thus possible, if the rapid
variation at the beginning of the observation is real rather than a
statistical fluctuation, that it originated in a young star rather
than the MBH.  The quiescent X-ray emission from these stars does not
exceed $L_{\rm X} \sim 10^{31}$ ergs s$^{-1}$, so a population of
$\sim 10^{2-3}$ magnetically active lower-mass stars would be needed
to produce all of the \sgrastar\ emission.  There are over a dozen
O-type stars in the central 0\farcs5 cusp.  If we were to use the
standard initial mass function (IMF) for stars in the solar
neighborhood, we would expect there to be about 100 magnetically
active low-mass stars per O star, so their combined luminosity would
be $L_{\rm X} \sim 10^{33-34}$ ergs s$^{-1}$.  However, it is believed
that the environment in the central parsec favors formation of
higher-mass stars and that the IMF in the central parsec may be
flatter and may have a higher low-mass cutoff than in the solar
neighborhood \citep{Morris93}.

The final stellar possibility that we consider for producing some
fraction of the X-rays is that of a population of compact stellar
objects in the entourage of the central black hole, \sgrastar\
\citep{Morris93,Lee95, Miralda-Escude00}.  If the massive stars now
observed in the central parsec evolve to produce stellar mass black
holes, and if those black holes are more massive than the bulk of the
field stars in the stellar population of the central stellar core,
then they will settle by gravitational segregation into a tight core
comparable in size to the observed stellar cusp.  If this process
occurs in a quasi-continuous fashion over the lifetime of the Galaxy,
then in the steady state, a substantial number of stellar-mass black
holes may be present in the compact central cluster, perhaps as many
as $10^{4-5}$.  Interestingly, the collective luminosity of such a
large number of compact objects, possibly including the most massive
neutron stars, cannot compete with emission by a single object of the
same total mass, because the Bondi accretion rate is proportional to
the square of the accretor mass \citep{Bondi52}.  If a black hole
cluster is to contribute substantially to the X-ray emission, then it
must contain close binaries with stellar companions that can
contribute a substantial accretion flow.

Whether or not X-ray binaries exist in the cusp is a topic of
considerable interest to stellar dynamicists.  The velocity dispersion
of stars in the cusp ($\ga 100$ km s$^{-1}$) is at least an order of
magnitude larger than in globular clusters ($\sim 10$ km s$^{-1}$), so
the favored mechanisms for forming binaries in globular clusters do
not work in the stellar cusp at the center of our Galaxy \citep[F.\
Rasio 2001, private communication; see also][]{Rasio93}.  Rough
estimates based on the tidal capture rate from hyperbolic orbits near
the MBH \citep[][; T.\ Alexander 2001, private
communication]{Alexander01} indicate that the number of X-ray binaries
in the cusp at any given time is at least 3 orders of magnitude less
than unity.  Furthermore, frequent collisions with surrounding stars
would cause any binaries to harden rapidly, leading either to
disruption of the main-sequence star or to formation of short-lived
common-envelope systems.  In either case, the lifetimes of binaries in
the cusp are probably relatively brief.

In summary, the X-ray luminosity and spectrum of \sgrastar\ are not
extremely different from those seen in colliding wind WR binaries, but
the absence of any \ion{He}{1}\ star coincident with the X-ray source
casts doubt on the presence of an appropriate binary at the correct
location.  The possible presence of one or more O+O binaries in the
central stellar cusp cannot be excluded, but it seems doubtful that
their spectra would be sufficiently hard.  A population of young,
magnetically active low-mass stars in the central stellar cusp could
produce the observed luminosity and spectrum, but there is no
observational evidence at this time that such low-mass stars are
actually present in the required numbers.  A population of $10^{4-5}$
compact stellar-mass objects accreting hydrodynamically from the
ambient medium could be present in the cusp, but their combined
luminosity would be many orders of magnitude fainter than that of a
single $\sim 10^6$ \msun\ black hole.  An origin in an accreting X-ray
binary system cannot be confidently excluded, but it seems improbable,
due to the difficulty of forming binaries in a stellar environment
with such a high velocity dispersion and to the rapid dynamical
evolution that would be expected for any binaries that might be
formed.  All things considered, a stellar origin for the emission
within 1\farcs5\ of \sgrastar\ is unlikely, and we proceed with the
discussion assuming the emission originates from accretion onto the
MBH.

\section{MBH Astrophysical Models of the X-ray Emission}
\label{sec:mbh_models}

Assuming the emission from the X-ray source we have detected with
\chandra\ is generated by matter accreting onto the massive black hole
associated with \sgrastar, we can use the measured X-ray luminosity
and spectrum to test the various black hole accretion models that have
been developed for \sgrastar.  As discussed in \S\ref{sec:intro}, one
would expect \sgrastar\ to emit $L_{\rm X} \sim3 \times 10^{43}$ ergs
s$^{-1}$ in the 2--10~keV band, if it were radiating at the Eddington
rate.  The observed 2--10~keV luminosity from \sgrastar\ reported in
this paper is $\sim 2\times 10^{33}$ ergs s$^{-1}$, which is
$\sim10^{10}$ times fainter than that.  It is possible that some
fraction of the observed emission could be contributed by stellar
objects within $\approx 1\arcsec$ of \sgrastar, so the actual ratio
could be even smaller.  \sgrastar\ is thought to accrete matter from
the stellar winds of nearby massive stars, particularly the dozen or
more luminous and windy \ion{He}{1}\ stars in the central parsec
cluster.  Current estimates for the Bondi capture rate range from
$\sim 3 \times 10^{-5}$ \msun\ yr$^{-1}$ \citep{Quataert99b} to $\sim
2 \times 10^{-4}$ \msun\ yr$^{-1}$ \citep{Coker97}.  Even at these
rates, \sgrastar\ is underluminous in X-rays, according to the
standard model, by factors of $\sim10^{7-8}$.

\subsection{Thermal Bremsstrahlung}
\label{sec:thermal_model}

\subsubsection{Spectral Shape}
\label{sec:shape}

The low luminosity of \sgrastar\ may be explained by accretion at a
rate much below the estimated Bondi rate, or by accretion at the Bondi
rate of gas that is radiating very inefficiently, or by some
combination of the two.  Since there appears to be an ample supply of
matter available from the stellar winds, research has concentrated on
the study of low radiative efficiency accretion flows.  Two prominent
models developed over the past decade have been the Bondi accretion
and the advection-dominated accretion flow (ADAF) models.  These
models assume quasi-spherical infall onto the MBH.

In the Bondi model \citep{Melia92,Melia94}, the highly supersonic
stellar winds flowing past \sgrastar\ form a bow shock that dissipates
the bulk motion of the gas and heats it to a temperature of
$\sim10^{7}$~K.  The ionized gas is then assumed to free-fall radially
with no net angular momentum until it reaches the circularization
radius at $R \sim 100\ R_{\rm S}$, where $R_{\rm S}$ is the
Schwarzschild radius of the MBH.  Gravitational binding energy
released during infall is transferred to the compressed magnetic
field, which heats the ionized gas through some combination of
magneto-sonic and/or magneto-turbulent processes.  Plasma
microinstabilities and collective effects are then invoked to set up
thermal equilibrium between the electrons and ions on a timescale
shorter than the infall timescale.  The ionized gas within the
Keplerian region is assumed to infall on a timescale much shorter than
the cooling time; consequently the thermal energy stored in the gas is
lost as the gas crosses the event horizon of the MBH.

In the ADAF models \citep{Ichimaru77,Rees82,Narayan95,Abramowicz95,
Mahadevan98}, the gas is assumed to accrete with angular momentum.
Turbulent magnetic viscosity dissipates energy and transfers angular
momentum outward through the flow, allowing the accreting material to
move inward.  The bulk of the viscous energy is assumed to be
deposited in the ions, with only a small fraction of the energy
transmitted directly to the electrons.  It is further assumed that
electrons and ions interact only via the Coulomb process.  The ions,
mainly protons, are unable to radiate efficiently and maintain a
temperature $T_{\rm i}$ close to the virial temperature at all radii
($T_{\rm i} \sim 10^{12}\ {\rm K}/r$, where $r = R/R_{\rm S}$).  The
electrons, on the other hand, radiate effectively via thermal
bremsstrahlung at larger radii and also via synchrotron and Compton
processes nearer the MBH.  At large radii, Coulomb scattering keeps
the electrons and ions at a common temperature, but at smaller radii
the relaxation timescale is longer than the infall timescale, and
their temperatures diverge.  The electron temperature begins to
saturate at $\sim 10^{3}$ R$_{\rm S}$, reaching a maximum of $\sim
10^{9-10}$~K near the MBH.  Thus, a two-temperature plasma develops in
the flow with the ions advecting the bulk of the released binding
energy through the event horizon.

The standard Bondi and ADAF models both assume the X-ray emission from
\sgrastar\ is dominated by thermal bremsstrahlung emission from
electrons in the hot optically thin accretion flow.  Both models can
fit the absorption-corrected 2--10~keV luminosity measured with
\chandra\ ($L_{\rm X} \simeq 2 \times 10^{33}$ ergs s$^{-1}$) by
adjusting the accretion rate downward by factors of a few from their
best-fit values to the \rosat\ upper-limit (but see
footnote~\ref{foot:rosat}\ in \S\ref{sec:x-ray}).  Given the large
uncertainty in the actual accretion rate onto \sgrastar\ from the
stellar winds, the measured luminosity alone cannot exclude either
model; however, it can be used to fix the accretion rate at the Bondi
radius and thus provides an important constraint on all hot accretion
flow models (see \S\ref{sec:mdot}).

For both models, the predicted intrinsic spectrum in the \chandra\
band (0.1--10~keV) has photon index $\Gamma \sim 1.4$ ($N(E) \propto
E^{-\Gamma}$) \citep{Melia94,Coker00,Narayan98a}.  However, in the
Bondi model the emission arises from a region within the
circularization radius ($\sim 100\ R_{\rm S}$), while in the ADAF
model the spectrum is dominated by emission from cooler gas at large
radii ($\ga 10^{4}\ R_{\rm S}$).  This difference in the location of
the dominant emitting region may be significant, as discussed below.

The best-fit, absorbed power-law model to the \chandra\ spectrum has
$\Gamma = 2.7^{+1.2}_{-1.0}$, which is much steeper than the predicted
Bondi/ADAF spectra in the literature.  Interestingly, the predicted
photon index for both models lies near the lower limit of the 90\%
confidence interval, even after adjusting for the systematic effects
discussed in \S\ref{sec:sgra_star_continuum}.  Due to the poor photon
statistics, we cannot exclude the Bondi/ADAF models with the current
data at the $3\sigma$ level.  However, it is possible that further
observations to measure the spectrum accurately might well enable us
to exclude them in the near future.\footnote{There is an additional
complication in comparing the observed spectrum to the ADAF model.
The ADAF spectrum is dominated by bremsstrahlung emission from the
outer parts of the accretion flow, so the spectral shape depends on
the relative size of the telescope beam with respect to the Bondi
accretion radius.  This can lead to a photon index steeper than 1.4
(E.~Quataert 2000, private communication).  Determination of the size
and radial profile of the extended X-ray emission from \sgrastar\ will
therefore be important when interpreting its spectrum.}  Either way,
the ability of \chandra\ to resolve the X-ray emission from \sgrastar\
out of the surrounding emission and to measure its spectrum will
provide information crucial to theoretical efforts to understand the
accretion and emission mechanisms in this source.

Recently, models have been developed in which it is proposed that not
all the matter in the accretion flow at large radii makes its way to
small radii and thence through the event horizon.  The Bernoulli
parameter in ADAF models is positive \citep{Narayan94,Blandford99},
which means the gas accretes with positive energy and may escape.
They propose that only a small fraction of the matter ($\sim 10^{-3}$)
in the outer accretion flow makes it to the event horizon and that the
binding energy released by this matter is transported to the outer
parts of the flow, where it drives a substantial wind that carries off
most of the matter.  Hence, the amount of matter actually accreting
through the event horizon is substantially reduced.  Lowering the
accretion rate at small radii would alleviate the need for the flow
onto \sgrastar\ to have a radiative efficiency $\eta \la 10^{-7}$.
This is a very appealing concept, since such a low radiative
efficiency would require that turbulent processes transfer energy from
the protons to the electrons at an extremely low rate.  As noted by
\citeauthor{Blandford99}, their adiabatic inflow-outflow solutions
(ADIOS) model generalizes the ADAF model to include the effects of a
wind.  \citet{Quataert99a}\ have computed spectral models for
\sgrastar\ using an ADAF+wind model.

It has also been realized that ADAFs are unstable to convection for
small values of the dimensionless viscosity parameter $\alpha \la
0.1$.  This has given rise to convection-dominated accretion flow
(CDAF) models \citep{Ball00,Quataert00a}, in which it is proposed that
convection transfers angular momentum inward and energy outward
through the flow.  The inward transfer of angular momentum almost
cancels out the normal outward transfer of angular momentum, and the
energy transported outward from the inner parts of the accretion flow
heats the outer regions of the flow, retarding the rate of accretion.
The net effect on the flow is that most of the matter circulates in
convective eddies at large radii, rather than falling inward toward
the event horizon.  Eventually, this excess matter must be lost from
the accretion flow in some manner, since otherwise a massive disk
would build up over time.

One consequence of these new models is that the mass density
distribution in the accretion flow rises less steeply toward the
center.  In the standard Bondi/ADAF models, the density ($\rho$)
varies with radius ($R$) as $\rho(R) \propto R^{-3/2}$, while in the
CDAF/ADAF+wind models $\rho \propto R^{-3/2+p}$.  Here $p$ is a
variable used to parameterize the effect of a wind or convection on
the density profile as a function of radius.  Consequently, the
accretion rate $\mdot$ is constant with radius in the Bondi/ADAF
models, while in the CDAF/ADAF+wind models it varies with radius
($\mdot(R) \propto R^{-p}$; $0 \leq p \leq 1$).

In the context of their ADAF+wind model, \citet{Quataert99a}\
investigate the effect of wind strength on the spectrum.  They find
that, ignoring the weak frequency dependence of the Gaunt factor, the
predicted spectrum in the \chandra\ band has photon index $\Gamma
\approx 3/2+2p/\epsilon$, where $\epsilon \approx 1$ is the power-law
index of the electron temperature profile ($T_{\rm e} \propto
R^{-\epsilon}$) in the outer parts of the flow.  Comparing this to our
measured spectrum of \sgrastar, we find that the ADAF model can be
reconciled with the soft spectrum ($\Gamma \sim 2.7$) of \sgrastar\ if
a substantial wind or strong convection ($p \approx 0.6$) is present.

Assuming that the observed X-ray spectrum is thermal bremsstrahlung
from a hot accretion flow onto \sgrastar\ and using the current
best-fit photon index, the predicted density profile through the flow
would vary roughly as $R^{-0.6}$ in these models.  Unfortunately, the
uncertainty in the measured photon index is too large to put tight
constraints on the parameter $p$ and hence on the density profile at
this time.  Values of $p$ ranging from $\approx 0$--1 are currently
acceptable.  Fortunately, the observations needed to put useful
constraints on the density profile are entirely feasible with
\chandra/ACIS.

\subsubsection{Accretion Rate and Emission Measure}
\label{sec:mdot}

Assuming the observed X-ray emission is from a hot, optically thin
thermal plasma accreting onto the MBH, we use our best-fit
Raymond-Smith model to estimate the accretion rate at the Bondi radius
($R_{\rm B}$) and to infer the accretion rate near the event horizon
for the various hot accretion flow models discussed above.  The
best-estimate ambient plasma conditions are $n_{\rm e} \approx 130$
cm$^{-3}$ and $kT_{\rm e} \approx 2$~keV; the corresponding emission
measure $EM \approx 2 \times 10^{3}$ cm$^{-6}$~pc, and the total mass
of the plasma is $\approx 2 \times 10^{-3}$~\msun.

The equation for the accretion radius is $R_{\rm B} = 2GM/c_{\rm
s}^2$, where $G$ is the gravitational constant, $M$ is the mass of the
black hole at \sgrastar, and $c_{\rm s}$ is the speed of sound in the
plasma.  The sound speed is given by the equation $c_{\rm s} = (\gamma
kT/\mu m_{\rm H})^{1/2} \approx 670$ km s$^{-1}$, which is comparable
to the bulk velocities of the stellar winds \citep[$\approx 200$--1000
km s$^{-1}$;][]{Paumard01}.  Here $\gamma$ is the adiabatic index, $k$
is Boltzmann's constant, $T$ and $\mu$ are the temperature and mean
atomic weight of the gas, and $m_{\rm H}$ is the mass of a hydrogen
atom.  For simplicity, we assume that the process is adiabatic
($\gamma = 5/3$) and that the gas is fully ionized with twice solar
abundances ($\mu \approx 0.70$).  Substituting the value for the sound
speed into the equation above, we find that $R_{\rm B} \approx 0.05\
{\rm pc}$ (1\farcs3), comparable to the 1\farcs5 radius of the circle
used to extract the spectrum.  In the analysis to follow, we adopt
$R_{\rm B} = 0.06$~pc (1\farcs5 or $2 \times 10^5\ R_{\rm S}$) for the
outer radius of the accretion flow.

Using a simple model by \citet{Bondi52}, the accretion rate at $R_{\rm
B}$ is then $\mdot_{\rm B} = 4\pi\lambda (GM)^2 \rho c_{\rm s}^{-3}
\simeq 3 \times 10^{-6} (n_{\rm e}/130\ {\rm cm}^{-3}) (kT/2\ {\rm
keV})^{-3/2}$ \msun\ yr$^{-1}$, where $\lambda = 1/4$ for an adiabatic
process, and $\rho = n_{\rm e} \mu m_{\rm H}$ is the plasma density.
This value for the accretion rate lies 1--2 orders of magnitude below
the most recent published estimates for the mass supply rate available
from the stellar winds (see the discussion at the beginning of
\S\ref{sec:mbh_models}).

The accretion rate for the standard ADAF model is related to the Bondi
accretion rate by $\mdot_{\rm ADAF} \sim \alpha \mdot_{\rm B}$, where
$\alpha$ is the dimensionless viscosity parameter in the standard thin
accretion disk model \citep{Shakura73}.  Then $\mdot_{\rm ADAF} \sim 3
\times 10^{-7} (\alpha/0.1) (n_{\rm e}/130\ {\rm cm}^{-3}) (kT/2\ {\rm
keV})^{-3/2}$ \msun\ yr$^{-1}$ at $R_{\rm B}$.  For the ADIOS/CDAF
models, the accretion rate scales as a function of radius as
$\mdot_{\rm ADIOS/CDAF} \sim \alpha \mdot_{\rm B} (R_{\rm S}/R_{\rm
B})^p$.  Using $p = 0.6$ derived above from the \chandra\ spectrum,
the predicted accretion rate across the event horizon would be
$\mdot_{\rm ADIOS} \sim 2 \times 10^{-10} (\alpha/0.1) (n_{\rm e}/130\
{\rm cm}^{-3}) (kT/2\ {\rm keV})^{-3/2}$ \msun\ yr$^{-1}$.  In the
CDAF model, the density profile is expected to be $\rho \propto
R^{-1/2}$, so $p$ should equal 1, which is within the uncertainties
allowed by the \chandra\ spectrum.  Thus, scaling the \chandra\
results using the CDAF model gives an accretion rate at the event
horizon of $\mdot_{\rm CDAF} \sim 1 \times 10^{-12} (\alpha/0.1)
(n_{\rm e}/130\ {\rm cm}^{-3}) (kT/2\ {\rm keV})^{-3/2}$ \msun\
yr$^{-1}$.

Recently, \citet{Aitken00}\ reported the detection of linearly
polarized radio emission from \sgrastar\ at frequencies above 150~GHz
(see \S\ref{sec:radio}).  \citet{Agol00}\ and \citet{Quataert00b}\
show that the detection of linear polarization implies the Faraday
rotation measure must be small and derive strong upper limits on the
density and magnetic field strength at small radii in the accretion
flow, where the polarized synchrotron emission is generated.  Both
groups find that $\mdot \la 10^{-8}$ \msun\ yr$^{-1}$ is required at
small radii to prevent Faraday rotation from depolarizing the
synchrotron emission and propose that a CDAF or ADAF with an outflow
could satisfy the conditions.  Comparing this limit derived from radio
observations to the estimates we derived above from our \chandra\ data
indicates that these models could plausibly satisfy this requirement.

Theoretical studies of X-ray emission lines in hot accretion flows
indicate that flows with strong winds should have stronger emission
lines than flows with weak or no winds \citep{Narayan99,Perna00}.
Detection of an Fe~K$\alpha$ line at 6.7~keV in the \sgrastar\
spectrum would be a strong argument for the existence of a hot thermal
accretion flow, since synchrotron models do not produce such a line,
and for the presence of a strong wind or convection.  In addition,
emission-line ratios could provide powerful diagnostics of the run of
density and temperature with radius in the flow.  The first crucial
step is to improve the signal-to-noise of the spectrum at 6--7~keV in
order to assess our possible detection of $K\alpha$ line emission from
highly ionized iron.  This will require significantly more observing
time with \chandra.

\subsection{Synchrotron Self-Compton}

Another emission mechanism which has been invoked recently to account
in particular for the X-ray flux from this source is inverse Compton
scattering of radio photons.  The population of relativistic electrons
responsible for producing the radio synchrotron emission is also held
responsible for upscattering those photons into the X-ray band by
inverse Compton scattering \citep{Beckert97}.  This hypothesis is
testable, inasmuch as it predicts that the X-ray flux is closely
linked to one component of the radio spectrum and that they should
vary in unison.

The primary distinction between existing synchrotron self-Compton
(SSC) models lies in the nature and location of the responsible
electrons.  \citet{Beckert97}\ invoke a quasi-monoenergetic electron
distribution at 30 or 40~$R_{\rm S}$ to account for the entire radio
spectrum and predict an SSC luminosity of $4 \times 10^{33}$ ergs
s$^{-1}$.  \citet{Falcke00}, on the other hand, associate the X-ray
flux with radio emission generated in the $\la 15\ R_{\rm S}$ nozzle
of a hypothetical jet.  In their model, the jet nozzle is responsible
for the sub-millimeter bump in the radio spectrum (the
centimeter-wavelength flux comes from much larger distances in the
jet), so it is the temporal variations of this component, on estimated
time scales of $\sim20$ days, which they expect to vary with the X-ray
flux.  A relativistic $\gamma$ factor for the electrons of about 100
is needed to produce both the correct sub-millimeter spectrum
\emph{and} the X-ray emission.

\citet{Melia00}\ similarly associate the X-ray flux with the
sub-millimeter bump via the SSC process.  They, however, place the
synchrotron source at $\sim 5\ R_{\rm S}$ in the inner Keplerian
region of the accretion disk, within the circularization radius of the
accreting plasma.  They stress the importance of simultaneity of
future radio and X-ray measurements, not simply because the radio
source is variable on time scales less than a year, but also because
the X-ray flux variations are expected to be much stronger than those
in the sub-millimeter.  A change of a factor of two in the
sub-millimeter flux would lead to a change in the X-ray flux by as
much as a factor or 10 or 20.

There are a number of measurements that could be used to distinguish
between the thermal bremsstrahlung and the SSC emission models.
First, the bremsstrahlung models do not predict the correlated
variability between the radio (cm to sub-mm) and the X-ray bands that
the SSC models do.  Second, X-rays from the SSC process would vary
rapidly as a result of the proximity of their source to the MBH, while
X-ray bremsstrahlung arising at larger radii should vary more slowly.
Third, the extended source we have observed should appear more
point-like when it brightens if some of the X-rays are produced via
the SSC mechanism near the MBH, while the more extended emission
requires a thermal mechanism.  Fourth, the X-ray spectrum should not
show strong emission lines if it is dominated by the SSC mechanism,
while the emission from the thermal model may show strong lines.  As
discussed above, it is possible that both emission components may be
present.  In that case, analyses of changes in the morphology and
spectrum of the source as it brightens and dims could be used to
determine the relative strengths of the two components.

\subsection{Role of the Local Diffuse X-ray Medium}
\label{sec:hot_ism}

The models discussed in the previous sections are based on the
assumption that the MBH is accreting matter from the stellar winds of
the nearby massive stars.  However, it is conceivable that the MBH is
embedded in a hot X-ray-emitting gaseous region and not the cooler
plasma ($\sim 10^4$~K) emerging from the stellar cluster.

It is evident from
Figures~\ref{fig:sgra_east}--\ref{fig:raw_central_arcmin} that hot
plasma is prevalent throughout the central $\sim 10$~pc of the Galaxy,
and is concentrated with higher densities in the innermost $\sim
1$~pc. This X-ray plasma will homogenize on timescales $l/c_{\rm s} <
10^3$ years, where $c_{\rm s} \simeq 560$ km s$^{-1}$ is the sound
speed and $l < 1$~pc is the characteristic length scale of the inner
region around \sgrastar.  The morphology of the diffuse emission
appears complex and is not symmetrical about \sgrastar\, suggesting
that it formed recently or is subject to external forces.

If the high surface brightness of the diffuse X-rays indicates the
presence of a local hot ISM surrounding \sgrastar, then it could be
the main source of material for accretion onto the MBH.  The local
diffuse plasma has $kT \simeq 1.3$~keV and density $n_{\rm e} \simeq
26$ cm$^{-3}$ (\S \ref{sec:diffuse_spec} and Table~\ref{tab:spec}). If
this plasma is stationary with respect to the MBH, the Bondi-Hoyle
accretion rate would be $\mdot_{\rm B, local} \sim 1 \times 10^{-6}$
\msun\ yr$^{-1}$ \citep{Bondi52}.  This accretion rate decreases if
the local medium is moving past the MBH at a rate $v_{\rm local} \geq
c_{\rm s}$, roughly as $\mdot_{\rm B} \propto v_{\rm local}^{-3}$.

The relationships between the X-ray emitting plasma seen in the ACIS
images and other gaseous structures observed or inferred to be present
are very unclear.  First, the stellar winds from the massive OB/WR
stars in the cluster centered $< 0.1$~pc from the MBH may create a
cavity of rapidly moving stellar gas within the hot ambient medium.
This is the model discussed above and in most studies of \sgrastar\
accretion.  The ram pressure of the winds is estimated to be 1--2
orders of magnitude greater than the thermal pressure of the ambient
medium and should dominate the ambient medium close to the stellar
cluster.  However, if the winds were to extend to very large distances
without thermalizing to X-ray temperatures, the observed cusp in the
diffuse X-ray emission should not be present.  The spatial
configuration of and the pressure balance between the cluster, MBH,
and ambient X-ray medium are thus not well established at the present
time.

The relationships between the ambient X-ray plasma and other gaseous
components in the Sgr~A region are similarly unclear.  First, as
discussed by \citet{Maeda01}, the local plasma may have been recently
compressed or pushed aside by the passage of the supernova shock wave
of Sgr~A East.  Second, the hot medium may envelop the cooler
spiral-shaped clouds of Sgr~A West (Fig.~\ref{fig:sgra_west}).  If
these clouds originated in the surrounding molecular ring, they are
likely subject to heating and evaporation as they orbit inward towards
the center.

At the present time, we are unable to determine whether the local
diffuse X-ray-emitting medium does or does not affect the accretion
onto the MBH.  However, future investigations of these issues must
consider the ambient hot plasma, which is imaged at high spatial
resolution for the first time in this study.

\section{Conclusions}
\label{sec:conclusions}

In this paper, we have presented results of our first-epoch
observation of the Galactic Center performed with the ACIS-I
instrument on the \emph{Chandra X-ray Observatory}.  We have produced
the first high spatial resolution ($\approx 1\arcsec$), hard X-ray
(0.5--7~keV) spectroscopic image of the central 40~pc (17\arcmin) of
the Galaxy.  Most importantly, we have resolved the X-ray emission
from the central parsec of the Galaxy and discovered a source,
\objectname[]{CXOGC J174540.0$-$290027}, coincident with the radio
position of \sgrastar\ to within 0\farcs35, corresponding to a maximum
projected distance of 16 light-days.  Over 150 point sources are
detected in the ACIS field of view: an increase in the X-ray source
density at the Galactic Center of more than an order of magnitude over
that detected by previous X-ray satellites.

A primary goal of this project is the search for an X-ray counterpart
to \sgrastar, the compact nonthermal radio source associated with the
MBH at the dynamical center of the Galaxy.  The X-ray source we have
detected at the position of \sgrastar\ has the following properties,
with the most secure results listed first:

\begin{enumerate}

\item The 2--10~keV luminosity is $L_{\rm X} \simeq 2 \times 10^{33}$
ergs s$^{-1}$, assuming the emission is isotropic
(\S\ref{sec:sgra_star_continuum}).  This is $\sim 10^2$ times fainter
than the upper limits obtained with previous X-ray satellites
(\S\ref{sec:x-ray}) and $\sim 10^{10}$ times fainter than the X-ray
luminosity that would be expected from the standard black-hole thin
accretion disk model, if the source were radiating at the Eddington
luminosity of the MBH (\S\ref{sec:intro}).  The extremely low observed
X-ray emissivity of the central MBH is a very powerful constraint on
any model.

\item The spectrum is well fit either by an absorbed power-law model
with photon index $\Gamma = 2.7$ or by an absorbed optically thin
thermal plasma model with $kT = 1.9$~keV.  In either case, the column
density $N_{\rm H} \simeq 1 \times 10^{23}$ cm$^{-2}$
(\S\ref{sec:sgra_star_continuum}).  The spectrum is softer than the
canonical AGN photon index with $\Gamma \simeq 1.5$--2.0.  This latter
result is not definitive due to the poor statistics of this faint
source.

\item The source appears extended with diameter $\approx 1\arcsec$
(\S\ref{sec:sgra_star_morph}).  This is very close to the resolution
limit of \chandra/ACIS and should be confirmed with higher photon
statistics from additional observations.

\item The inner region of this small structure may have varied on
timescales of $\simeq 1$~hr with a factor of 2 amplitude
(\S\ref{sec:sgra_star_ltc}).  This result is also photon limited, with
$2.7\sigma$ significance.

\item Tentative evidence for a $\rm K\alpha$ line at 6.7~keV from
He-like iron is seen with $\simeq 2\sigma$ significance
(\S\ref{sec:sgra_star_fek}).

\end{enumerate}

Based on the fluxes and the spatial distribution of the X-ray sources
in the field, we estimate that the probability of detecting an
absorbed source by random chance that is as bright or brighter than
the \sgrastar\ candidate and that lies within $0\farcs35$ of the radio
position is $P \la 6 \times 10^{-3}$.  The nearest windy \ion{He}{1}\
emission-line stars are too far away ($\geq 1\farcs$2) to account for
the X-ray source, and other classes of normal stars are too soft to
penetrate the high absorbing column.  Colliding winds in O+O binaries
in the central stellar cusp may be able to produce the required
luminosity, but their spectra might be too soft.  A large population
of young, magnetically active low-mass stars in the cusp could produce
the observed luminosity and spectrum, but there is currently no
evidence in the radio or IR bands that such stars are actually
present.  A population of $10^{4-5}$ compact stellar-mass objects
accreting hydrodynamically from the ambient medium appears an unlikely
origin for the emission, since their cumulative luminosity would be
many orders of magnitude fainter than that of a single $\sim 10^6$
\msun\ black hole.  It is possible that the emission could originate
in an accreting X-ray binary system within the cusp, but the expected
number of X-ray binaries is $< 10^{-3}$.

Assuming the observed emission within 1\farcs5 of \sgrastar\ is from
accretion onto the MBH, we can use the observed properties of the
source to constrain the models.  Due to the limited photon statistics,
the luminosity and spectral shape of the source can be fit either by
an optically thin thermal plasma model or by an SSC model.  The
apparent extent of the source and the possible detection of an Fe line
support the thermal model, while the possible detection of rapid
variability supports the SSC model.  The current observations, while
of limited signal-to-noise, are thus consistent with the presence of
both thermal and nonthermal emission components in the \sgrastar\
spectrum.

The results presented in this paper have demonstrated the great
potential of \chandra/ACIS to revolutionize our understanding of
highly energetic phenomena in the central parsec of our Galaxy.  No
other X-ray satellite for the foreseeable future will have its unique
combination of arcsecond resolution, high sensitivity, broad energy
band, and moderate spectral resolution.  These properties are
indispensable for this study.  The results also show that further
observations are needed to increase the photon statistics.  An
improved spectrum could be used to constrain the continuum shape and
to search for an emission line at 6.7~keV.  Detection of such a line,
for example, would show conclusively that a thermal component is
present, while a nondetection would put tight constraints on the
strength of a wind or outflow in the thermal models.  The SSC models
predict that the X-ray emission should be variable and show a close
correlation with variations in the millimeter band.  A vigorous
campaign of simultaneous monitoring in the X-ray and millimeter bands
is needed to test this prediction.  Likewise, increased photon
statistics could be used to measure changes in the morphology of the
emission from the source, if it varies, and to examine the size of the
source as a function of energy.  An order of magnitude longer exposure
will be needed to achieve these goals.

\acknowledgments

We thank Farhad Yusef-Zadeh for allowing us to use his VLA 6-cm image
of \sgrastar, Peter Predehl for discussions regarding the \rosat\
observation of \sgrastar, Heino Falcke and Fulvio Melia for sharing
with us the results of their latest models prior to publication, Eliot
Quataert for discussions on the similarities and differences between
various hot accretion flow models, and Rashid Sunyaev for sharing with
us his physical insights into accretion flows and X-ray sources in the
Galactic Center.  Finally, we thank all the members of the ACIS
instrument team at MIT and Penn State and all the people at the CXC,
TRW, Ball Aerospace, Hughes-Danbury, and Eastman/Kodak who have worked
so long and hard to build, launch, and operate this great X-ray
observatory in space, without which the observations reported here
could not have been performed.  This research was supported by NASA
grant NAS 8-38252.  W.~N.~Brandt acknowledges support from NSF CAREER
award AST-9983783.

\clearpage
\newpage

\begin{deluxetable}{lcccccl}
\tabletypesize{\scriptsize}
\tablewidth{0pt}
\tablecaption{X-ray Positions and Count Rates of \tycho\ Astrometric
  Sources\label{tab:tycho_astrom}}
\tablehead{
\colhead{CXOGC Name} &
\colhead{R.A.\tablenotemark{a}} &
\colhead{Dec.\tablenotemark{a}} &
\colhead{$\Delta$R.A.\tablenotemark{b}} &
\colhead{$\Delta$Dec.\tablenotemark{b}} &
\colhead{Count Rate\tablenotemark{c}} &
\colhead{\tycho\,\tablenotemark{d}} \\
\colhead{} &
\colhead{(J2000)} &
\colhead{(J2000)} &
\colhead{(\arcsec)} &
\colhead{(\arcsec)} &
\colhead{($\times 10^{-3}$ counts s$^{-1}$)} &
\colhead{}}
\startdata
J174525.7$-$285626 & 17 45 25.78 $\pm$ 0.01 & $-$28 56 26.8 $\pm$ 0.2 & \phs0.0 & $-$0.1 & $\phn1.54\pm0.20$    & 6840-666-1\tablenotemark{e} \\
J174530.0$-$290704 & 17 45 30.01 $\pm$ 0.01 & $-$29 07 04.4 $\pm$ 0.2 &  $-$0.1 & $+$0.2 & $11.3\phn\pm0.5\phn$ & 6840-020-1\tablenotemark{f} \\
J174543.9$-$290456 & 17 45 43.92 $\pm$ 0.01 & $-$29 04 56.4 $\pm$ 0.2 &  $-$0.1 & $-$0.2 & $\phn3.65\pm0.30$    & 6840-590-1 \\
\enddata
\tablenotetext{a}{Positional uncertainties are derived by combining
the statistical uncertainties from centroiding with the residual RMS
scatter in the X-ray positions of the \tycho\ reference sources after
aligning \chandra\ to the \hipparcos\ celestial coordinate system
(\S\ref{sec:astrom}).}
\tablenotetext{b}{Positional offsets are defined as \tycho\ position
\citep{Hog00} $-$ \chandra\ position.}
\tablenotetext{c}{Count rate in the 0.5--1.5~keV band.}
\tablenotetext{d}{\tycho\ identifiers are from \citealt{Hog00}.}
\tablenotetext{e}{Optical identifier: \objectname[]{CSI-28-17423} (B
star).}
\tablenotetext{f}{Optical identifiers: \objectname[]{HD~316224} (F2
star); \objectname[]{HIP~86911} (parallax = 10.73~mas).}
\tablecomments{Units of right ascension (R.A.) are hours, minutes, and
seconds; units of declination (Dec.) are degrees, arcminutes, and
arcseconds.}
\end{deluxetable}

\clearpage
\newpage

\begin{deluxetable}{lccccc}
\tabletypesize{\scriptsize}
\tablewidth{0pt}
\tablecaption{X-ray Position and Count Rate of
  \sgrastar\label{tab:sgra_star_astrom}}
\tablehead{
\colhead{CXOGC Name} &
\colhead{R.A.\tablenotemark{a}} &
\colhead{Dec.\tablenotemark{a}} &
\colhead{$\Delta$R.A.\tablenotemark{b}} &
\colhead{$\Delta$Dec.\tablenotemark{b}} &
\colhead{Count Rate\tablenotemark{c}} \\
\colhead{} &
\colhead{(J2000)} &
\colhead{(J2000)} &
\colhead{(\arcsec)} &
\colhead{(\arcsec)} &
\colhead{($\times 10^{-3}$ counts s$^{-1}$)}}
\startdata
J174540.0$-$290027 & 17 45 40.02 $\pm$ 0.01 & $-$29 00 27.8 $\pm$ 0.2 & $+$0.2 & $-$0.2 & $5.51\pm0.42$ \\
\enddata
\tablenotetext{a}{Positional uncertainties are derived by combining
the statistical uncertainties from centroiding with the residual RMS
scatter in the X-ray positions of the \tycho\ reference sources after
aligning \chandra\ to the \hipparcos\ celestial coordinate system
(\S\ref{sec:astrom}).}
\tablenotetext{b}{Positional offsets are defined as radio position
\citep{Yusef-Zadeh99} $-$ \chandra\ position.}
\tablenotetext{c}{Count rate in the 0.5--9.0~keV band.}
\tablecomments{Units of right ascension (R.A.) are hours, minutes, and
seconds; units of declination (Dec.) are degrees, arcminutes, and
arcseconds.}
\end{deluxetable}

\clearpage
\newpage

\begin{deluxetable}{llccc}
\tabletypesize{\footnotesize}
\tablewidth{0pt}
\tablecolumns{5}
\tablecaption{Spectral Fits to X-ray Sources in the Central 10\arcsec\
    of the Galaxy\tablenotemark{a} \label{tab:spec}}
\tablehead{
\colhead{} &
\colhead{} &
\colhead{\sgrastar} &
\colhead{Integrated Point\tablenotemark{b}} &
\colhead{Local Diffuse} \\
\colhead{} &
\colhead{} &
\colhead{} &
\colhead{Sources} &
\colhead{Emission}
}
\startdata
\multicolumn{5}{l}{Power-law Model} \\
& $\rm N_H$ [$\times 10^{22}$ cm$^{-2}$] & 9.8~(6.8--14.2) & 13.2~(10.5--19.7) & \nodata \\
& $\Gamma$ & 2.7~(1.8--4.0) & 2.5~(1.9--3.7) & \nodata \\
& Norm.\ [$\times 10^{-4}$ photons cm$^{-2}$ s$^{-1}$ keV$^{-1}$ at 1~keV] & 3.5~(0.6--28.8) & 8.2~(2.8--75.0) & \nodata \\
& $\rm \chi^2/d.o.f.$ & 19.8/22 & 56.3/72 & \nodata \\
\multicolumn{5}{l}{Optically Thin Thermal Plasma Model\tablenotemark{c}} \\
& $\rm N_H$ [$\times 10^{22}$ cm$^{-2}$] & 11.5~(8.4--15.9) & \nodata & 12.8~(11.4--14.2) \\
& kT [keV] & 1.9~(1.4--2.8) & \nodata & 1.3~(1.2--1.5) \\
& Norm.\ [$\times 10^{-4}$ cm$^{-5}$]\tablenotemark{d} & 5.2~(2.7--12.3) & \nodata & 62~(43--87) \\
& $\rm \chi^2/d.o.f.$ & 16.5/22 & \nodata & 119.1/121 \\
\multicolumn{5}{l}{Flux and Luminosity (2--10~keV)\tablenotemark{e}}
\\ & $\rm F_X$ [$\times 10^{-13}$ ergs cm$^{-2}$ s$^{-1}$] &
1.3~(1.1--1.7) & 4.0~(3.3--4.5) & 6.0~(5.6--6.2)\tablenotemark{f} \\ &
$\rm L_X$ [$\times 10^{33}$ ergs s$^{-1}$] & 2.4~(1.8--5.4) &
7.8~(6.0--19.4) & 24~(18--32)\tablenotemark{f} \\ \enddata
\tablenotetext{a}{See \S\ref{sec:sgra_star_continuum},
\S\ref{sec:pt_src_spec}, and \S\ref{sec:diffuse_spec} for details.}
\tablenotetext{b}{Counts from \sgrastar\ are excluded from the
integrated spectrum of the point sources within 10\arcsec\ of
\sgrastar.}
\tablenotetext{c}{The optically thin thermal plasma code that we use
was developed by \citet{Raymond77}.  Twice solar elemental abundances
are assumed.}
\tablenotetext{d}{${\rm Normalization} = 10^{-14} \int n_{\rm e}
n_{\rm i} dV / 4\pi D^2$, where $n_{\rm e}$ and $n_{\rm i}$ are the
electron and ion densities (cm$^{-3}$) and $D$ is the distance to the
source (cm).}
\tablenotetext{e}{X-ray luminosities ($L_{\rm X}$) are corrected for
absorption, while fluxes ($F_{\rm X}$) are not corrected.  The
quantities $F_{\rm X}$ and $L_{\rm X}$ listed for \sgrastar\ are
derived using the best-fit parameters of the power-law model; the
optically thin thermal plasma model gives similar values.}
\tablenotetext{f}{Divide by $\pi \times 10^{2}$ square arcseconds to
convert to surface brightness.}
\tablecomments{The uncertainties on the model parameters are the 90\%
confidence intervals computed using $\Delta\chi^2 = 2.71$ for one
interesting parameter.  The uncertainties for the fluxes and
luminosities are the 90\% confidence intervals computed using
$\Delta\chi^2 = 4.61$ for two interesting parameters
(\S\ref{sec:sgra_star_continuum}).}
\end{deluxetable}

\clearpage
\newpage

\newpage

\newpage

\newpage

\newpage

\newpage

\newpage

\newpage

\newpage

\newpage

\end{document}